\documentclass[traditabstract]{aa} 
\usepackage{graphicx}
\usepackage{txfonts}
\usepackage{aalongtable}
\usepackage{afterpage}
\begin{document}

\def\ho{H$_0$}
\def\kms{km\,s$^{-1}$}
   \title{The DPOSS II distant compact group survey: the EMMI-NTT spectroscopic 
sample\thanks{Based on observations obtained from programs 074.A-0460, 075.A-0520, 279.A-5048}}

   \subtitle{}

   \author{E. Pompei\inst{1}$^{,}$\inst{2} \and A. Iovino\inst{2}}
         
   \offprints{Emanuela Pompei: epompei@eso.org}

   \institute{European Southern Observatory,
Alonso de Cordova 3107, Vitacura, Casilla 19001, Santiago 19, Chile
	\and
Osservatorio Astronomico di Brera
Via Brera 28, I-20121 Milano, Italy;\newline
emanuela.pompei@brera.inaf.it;
iovino@brera.mi.astro.it}

   \date{Received September 30, 2011/ Accepted:}

 
  \abstract
   {We present the results of a redshift survey of 138 candidate
compact groups from the DPOSS II catalogue, which extends the available redshift range of
spectroscopically confirmed compact groups of
galaxies to redshift $z \sim 0.2$.\newline
In this survey we aim to confirm group membership via spectroscopic
redshift information, to measure the characteristic properties of
the confirmed groups, namely their mass,
radius, luminosity, velocity dispersion, and crossing time, and to compare them 
with those of nearby compact groups. Using 
information available from the literature, we also studied the surrounding group 
environment and searched for additional, previously unknown, group 
members, or larger scale structures to whom the group might be
associated. Among the 138 observed groups, 96 had 3 or more concordant galaxies,
i.e. a 70$\%$ success rate. Of these 96, 62 are isolated on the sky, while
the remaining 34 are close on the sky to a larger scale structure. 
The groups which were not spectroscopically confirmed as such turned out to be couples of pairs
or chance projections of galaxies on the sky.

The median redshift of all the confirmed groups is $z \sim 0.12$,
which should be
compared with the median redshift of 0.03 for the local sample of
Hickson compact groups. 
The average group radius is 50 Kpc, and the median radial velocity
dispersion is 273\kms, while typical crossing times range from 
0.002 H$_{0}^{-1}$ to 0.135 H$_{0}^{-1}$ with a median value of 
0.02 H$_{0}^{-1}$, which are quite similar to the values measured for the Hickson
compact group sample.
The average mass-to-light ratio of the whole sample, M/L$_{B}$, is 
192, which is significantly higher than the value measured for Hickson's compact 
groups, while the median mass, measured using the virial theorem, is 
M = 1.67$\times$ 10$^{13}$M$\odot$. When we select only the groups that
are isolated on the sky the values quoted become lower and closely resemble the
average
values measured for Hickson's compact groups. We also find that the characteristics of the groups
depend on their environment.

We conclude that we observe a population of compact
groups that are very similar to those observed at zero redshift. Furthermore, a
careful selection of the environment surrounding the compact groups is
necessary to detect truly isolated compact structures.}


\titlerunning{The EMMI-NTT compact groups survey}

    \maketitle

%

\section{Introduction}

Compact groups (hereafter CGs) of galaxies are small associations of galaxies on the sky
characterized by a few members, of the order of four to eight, by a relatively
small velocity dispersion, of the order of 200\kms, that are separated on the sky
by an average distance comparable to the diameter of individual galaxies.
Under the assumption that CGs are gravitationally
bound objects free from other external influences, one expects these
CGs, owing to their mutual interactions, to evolve rapidly, following 
several violent interactions among the member galaxies, 
and to form a single isolated early type galaxy in a  short time interval
compared with the Hubble time (see for example Barnes, 1989).\newline
Nevertheless, despite many multiwavelength observations and intensive
analyses have
confirmed that many compact groups are gravitationally bound objects (e.g. 
Verdes-Montenegro et al. 2001 (VM01), Ponman et al., 1996; Mendes de Oliveira et al., 
1994 (MDO); Hickson et al., 1992), the picture that has emerged is remarkably more 
complex than the one suggested by earlier studies. To begin with,
members of isolated compact
groups had only a small fraction of strongly interacting galaxies, of
the order of 7$\%$ (MDO, 94), in contrast to the expectations of N-body simulations.
However, evidence of gentler interactions (e.g. gas stripping) was detected
in almost half of the member galaxies, implying that there was some
kind of influence of the group environment on the evolution of its members. 
Other correlations, e.g. the one between velocity dispersion and
dominant morphological type in groups,
and that between crossing time and spiral fraction (Hickson et al.,
1992) implied that CGs  follow an evolutionary path that
leads 
to several possible endings. Compact
groups were then proposed to merge and to form an isolated early type galaxy, or,
depending on the original mass, a fossil group. Alternatively,
it was proposed that their lifetimes were much longer than 
predicted by early numerical simulations owing to
a massive halo of dark matter stabilizing the compact group for a long
time.
This excluded a short lifetime and explained the
difficulty in identifying the final merging product of a CG.\newline
Not all compact groups were, however, found to be as isolated on the sky as originally
supposed (see for example de Carvalho et al., 1994; Ribeiro et al.,
1998). Some of them were found to be quite close to clusters of
galaxies whose richness seems to vary with redshift (Andernach \&
Coziol, 2005),
while the remainder could be divided into three categories.
These were named: (1) {\it loose groups}, i.e. a larger
than previously expected galaxy concentration; (2) a {\it core+halo}
configuration, i.e. a central
concentration within a looser distribution of galaxies; and finally, (3) compact groups that
complied with the original definitions, i.e were truly isolated and
gravitationally bound dense
structures. This led many authors to propose that CGs are a
local universe phenomenon in a biased cold-dark-matter galaxy
formation model (West, 1989, Andernach \& Coziol 2005, to cite a
few). Larger scale structures would then form first, leaving smaller
associations such as compact
groups to form last, with or shortly before field galaxies.
The different kind of groups observed were just different structures
at different spatial scales and formation times.\newline
In constrast, Einasto et al., 2003, showed that loose groups of
galaxies close to large-scale structures are on average more massive and have
a larger velocity dispersion than those that are more isolated on
the sky. According to these authors, this is evidence that the
large-scale 
gravitational field responsible for the formation of rich
clusters enhances the evolution of neighbouring poor systems.
A larger velocity dispersion implies a higher mass, i.e. that an
environmental enhancement of mass is observed. This in turn is
interpreted as direct evidence of the hierarchical formation of galaxies and
clusters in a network of filaments connecting high density knots of the
cosmic mass density.\newline

To shed some light on the evolutionary
path of CGs and on their relation to environment, and in
order to understand what is the role of CGs in the evolution of their
member galaxies
and the larger-scale structures, it is
imperative to extend to higher redshift the available samples and to
conduct a detailed study of the surroundings of the observed compact groups.

A dedicated search for more distant (up to z$\sim$0.2) CGs
was started in earnest five to six years ago, with the compilation of a catalogue describing a
pilot sample of distant groups drawn from the second digital
Palomar Observatory Sky Survey (hereafter DPOSS II) of Iovino et al.,
2003, which was later complemented by a catalogue of GCs from SDSS early
release (Lee et al., 2004), and yet more complete catalogues (de Carvalho et al.,
2005; McConnachie et al., 2009, Tago et al., 2010). Most of the
information contained in these catalogues is based on photometric data,
while spectroscopic redshift information is available for at most two
galaxies in each group. Despite this, the main results that
could be drawn from these studies were that distant CGs were very 
similar to their nearby counterparts. We note, however, that the 
group environment was never considered in any of these aforementioned studies.\newline
Until now, no detailed spectroscopic follow-up has been performed for 
any distant CGs sample, with the exception of two pilot studies,
one by Pompei et al, 2006, for a small sample of DPOSS II CGs, and
another by Gutierrez, 2011 for three CGs at z$\sim$ 0.3 drawn from the
catalogue of McConnachie et al. (2009). This has so far
limited any deeper study of the properties of CGs outside our
neighbourhood. To obviate this lack, we started a spectroscopic
follow-up campaign on the DPOSS II compact group sample, from 2004
to 2008, observing each member galaxy in 138 candidate compact groups from the DPOSS II
catalogue, which was the only large catalogue of distant CGs
available when our observations started.\newline

We present in this paper our
main results for the whole sample of 138 CGs candidates, deferring to a future paper the
discussion of the spectroscopic properties of the member galaxies,
the percentage of active galactic nuclei, and the presence of anemic spirals in
compact groups.\newline
The paper is organized as follows: in Section 2, we describe our
observations and data reduction, in Section 3 we present our results,
and in Section 4 we discuss the possible implications for the
evolution
of CGs. In Section 5 we provide our conclusions.

\section{The data}

The sample was selected from the DPOSS II compact group catalogue
(Iovino et al. \cite{iovino03},
de Carvalho et al.\cite{reina05}) depending on the allocated observing windows.
The most comprehensive coverage was between 09 $\le$ RA $\le$17h and -1$^{o} \le$
DEC $\le +15^{o}$, but a
few candidates at other 
coordinates were also observed. This sample is representative of
the DPOSS CGs catalogue, but is by no means complete in either magnitude
or redshift.

\subsection{Observations and data reduction}
The observations and data reduction were carried out in the same way
as described in Pompei et al., 2006, hereafter Paper I, and we describe
them here briefly for completeness sake.\newline
All the data were obtained with the 3.58m New Technology Telescope (NTT)
and the ESO Multi Mode Instrument (hereafter EMMI) in
spectroscopic mode in the red arm, equipped with
grism $\#$2 and a slit of 1.5$\arcsec$, under clear/thin cirrus
conditions and grey time. The MIT/LL red arm detector, a mosaic
of two CCDs 2048 x 4096,  was binned by two
in both the spatial and spectral directions, with a resulting dispersion of
3.56$\AA$/pix, a spatial scale of 0.33$\arcsec$/pix, an instrumental
resolution of 322 \kms, and a wavelength coverage from 3800 $\AA$ to
9200 $\AA$. When possible, two or more galaxies were placed together
in the slit, whose position angle had been constrained by the location
of galaxies in the sky and thus almost never coincided with the
parallactic angle. Exposure times varied from 720s to 1200s per spectrum, and
two spectra were taken for each galaxy to ensure reliable
cosmic ray subtraction. When the weather conditions allowed it, spectrophotometric
standard stars were observed during the night, to flux calibrate the
reduced spectra.
Standard data reduction was performed using the MIDAS data reduction package\footnote{Munich
Image Data Analysis System, which is developed and maintained by the
European Southern Observatory} and our own scripts. Wavelength calibration
was applied to the two-dimensional (2D) spectra and an upper limit of 0.16$\AA$
was found for the rms of the wavelength solution.\newline 
The two one-dimensional spectra available for each galaxy were averaged
together at the end of the reduction, giving an average
signal-to-noise ratio (S/N) of $\sim$
30 (grey time) or $\sim$ 10-15 (almost full moon) per resolution
element at 6000$\AA$.\newline
Flux calibration was possible for 80$\%$ of the nights, with an average error
of 10-15$\%$.  For the other nights, the conditions were too variable to
obtain a reliable calibration.\newline
Radial velocity standards from the Andersen et al. (1985) paper were
observed with the same instrumental set-up used for the target
galaxies; in addition to this, we also used galaxy templates with
known spectral characteristics and heliocentric velocity available
from the literature, i.e.  M32, NGC 7507, and NGC 4111.

\subsection{Redshift measurement}
The technique is similar to the one used in Paper I, and again we briefly 
mention the points that are most important to understand the current
data set.\newline
The IRAF\footnote{IRAF is distributed by NOAO, which is operated by AURA,
Inc, under cooperative agreement with the NSF} packages {\it xcsao}
and {\it emsao} were used to
measure the galaxy redshifts by means of a cross-correlation method
(Tonry \& Davis, 1979), where robust measurements were obtained for galaxy spectra
dominated by either emission lines or absorption lines.  For spectra
dominated by absorption lines, we used galaxy templates and stellar
radial velocity standards, while for emission-line dominated spectra we
used a synthetic template generated by the IRAF package {\it
linespec}.  Starting from a list of the stronger emission lines
(H$\beta$, [OIII], [OI], H$\alpha$, [NII], [SII]), the package creates
a synthetic spectrum, which was then convolved with the instrumental
resolution.\newline
A confidence parameter (see Kurtz \& Mink (1998) for a complete
discussion) was used to assess the goodness of the estimated
redshift: all redshifts with a confidence parameter {\it r}
$\ge$ 5 were considered reliable, while measurements with 2.5 $\le$ {\it r} $\le$ 5 were
checked by hand.  Measurements with {\it r} $\le$ 2.5 are not
reliable. All the confirmed member galaxies in our sample had
a {\it r} $>$ 3.5.

In some cases, {\it emsao} failed to correctly identify the emission
lines, which happened each time some emission lines were contaminated
by underlying absorption.  When this occurred, we measured the
redshift by gaussian fitting the strongest emission lines visible
and took the average of the results obtained from each line.
If two or more lines were blended, the IRAF command {\it deblend}
within the {\it splot} package was used.\newline

The recession velocity errors varied between 15 and 100 \kms, depending 
on the kind of the galaxy spectrum (emission or absorption dominated)
and also the S/N of the target spectrum.\newline

Corrections to produce heliocentric recession velocities were
estimated using
the IRAF task {\it rvcorrect} in the package {\it noao.rv}.\newline

Whenever possible, we checked the existing literature for other published
redshifts; in particular, we made extensive use of the overlapping area with the SDSS,
SDSS-R7 (see Abazajian et al., \cite{aba09}). For galaxies that were well separated from other
objects, we found a remarkable agreement with exisiting measurements within
our measurements errors (see Fig. 1). 

\setcounter{figure}{0}
\begin{figure*}
\includegraphics[scale=0.5]{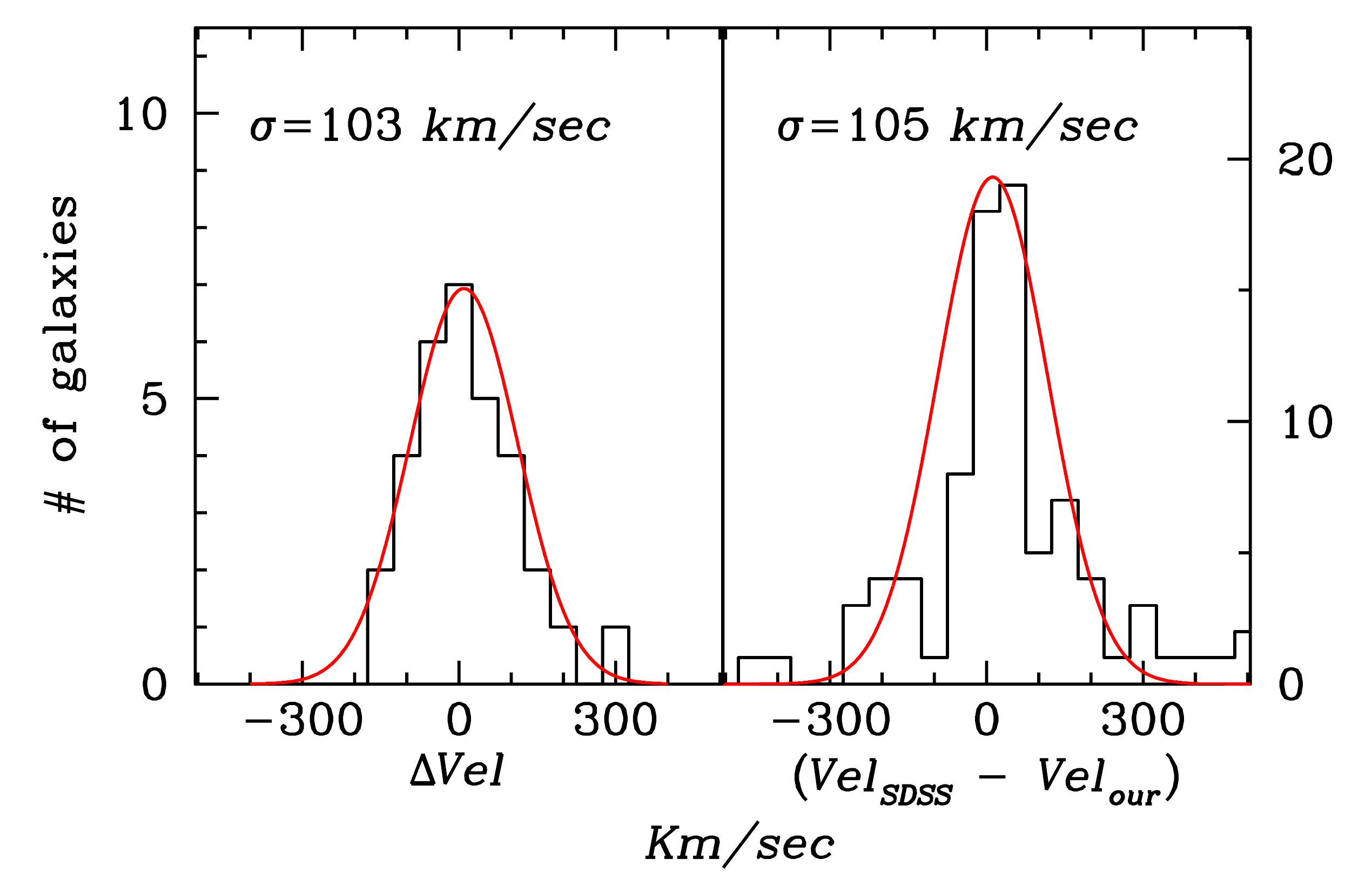}
\caption{Left-hand panel: distribution of our measured velocity error.
Right-hand panel: comparison between our measured radial velocities
and
those from the SDSS: the agreement between the two data sets is very
good within the errors.}
\label{fig:SDSS_velo}
\end{figure*}

Some disagreements were found instead for
galaxies close on the sky to either nearby stars or other galaxies with
different redshifts. We assumed that this happened mostly because of
the fiber proximity limit, which prohibited the Sloan spectrograph from
observing targets closer to each other on the sky than 60$\arcsec$ at
z=0.1
in a single pass.
We note that when the only available redshifts were photometric, these
frequently disagreed with our own spectroscopic measurements.

\subsection{Luminosity measurements}
We briefly summarize here how we estimated the luminosity of
each member galaxy; for a full description, the reader can consult paper I.
The luminosity of each group was obtained by summing up all the luminosities
of the member galaxies, after correction for Galactic extinction and k-correction.
Only two values of {\it k} correction were used, one for early-type galaxies (E-Sa),
and another for late-type ones (Sb onward), identified by an
EW(H$\alpha$) $>$ 6$\AA$\footnote{We use throughout the paper a
  positive sign for emission lines and a negative one for absorption lines.} 
and morphological evidence, i.e. presence of spiral arms.
No correction for passive evolution was applied, both because not all galaxies in
our sample can be characterized by a simple stellar population and
owing to the
large r.m.s which is comparable at z=0.1 to the amount of correction that
would be applied, assuming the models from van Dokkum et al. no. 1, 2, and 3 (Longair,
2008).
All R band luminosities were converted to B band luminosity using the transformation 
(Windhorst et al., 1991) based on the empirical relations of Kent (1985):

\begin{equation}
B  =  g+0.51+0.60\times(g-r)
\end{equation}

and assuming that $M_\mathrm{B,\sun}$ = 5.48.

Errors in the luminosity were estimated by assuming the maximum error
in the photometric calibration of DPOSS plates, i.e. an error of 0.19
magnitudes for an {\it r} magnitude of 19 (Gal et al. 2004).

\section{Results}
This section is divided into two parts. The first part will
describe our so-called sample cleaning, i.e. an analysis
of the environment surrounding the spectroscopically confirmed groups,
to determine whether they fulfil the isolation criteria.
This basically divides the confirmed groups into two categories: (1) the objects really isolated
on the sky, which can be assumed to be {\it bona fide} compact groups;
and, (2) objects close on the sky to larger scale structure, to whom they may
be associated, or objects that {\bf are} part of a cluster of galaxies
and have been selected as candidate CGs by mistake.\newline

In the second part of this section, we calculate the so-called characteristic
parameters of the CGs, namely velocity dispersion, crossing time, radius,
and mass, using various estimators, and we compare our measurements with other
existing works and with the values of the same parameters obtained for compact groups in the
nearby universe.

\subsection{Group membership and environment}
We consider a candidate compact group to be spectroscopically confirmed if
at least three of its members have accordant redshifts, i.e. are within $\pm$
1000\kms of the median redshift of the group.
The redshift of the confirmed group is assumed to be the median value of the
measured redshift of its confirmed members.\newline
To calculate the median group velocity and its radial velocity
dispersion, we used the 
biweight estimators of location and scale (Beers et al.,
1990). Among 138 candidate groups, we confirmed 96 concordant objects, i.e. $\sim$ 70$\%$
success rate.

Our measured recession velocities, {\it cz}, range from 13263\kms to 67724\kms
with an average value of 34792\kms, i.e z = 0.116, one order of magnitude larger
than the average value for nearby compact group catalogues.
The velocity distribution of our confirmed compact groups is shown in
Figure 2.

We then proceeded to search the environment surrounding each group using available 
catalogues from the literature and the last SDSS release, and adopting a search radius 
equal to the Abell radius of a cluster at the distance of each 
redshifted  compact group.

Whenever redshift measurements were available for the literature
catalogues used in our investigation of the environment, we refined our search, by assuming that the
group is close to a cluster with which it may be associated, if the redshift difference between the two was $\Delta$z $<$ 0.01,
i.e. a velocity difference of 3000\kms, one order of magnitude above the typical velocity
dispersion of compact groups.

\setcounter{figure}{1}
\begin{figure}
\includegraphics[scale=0.5]{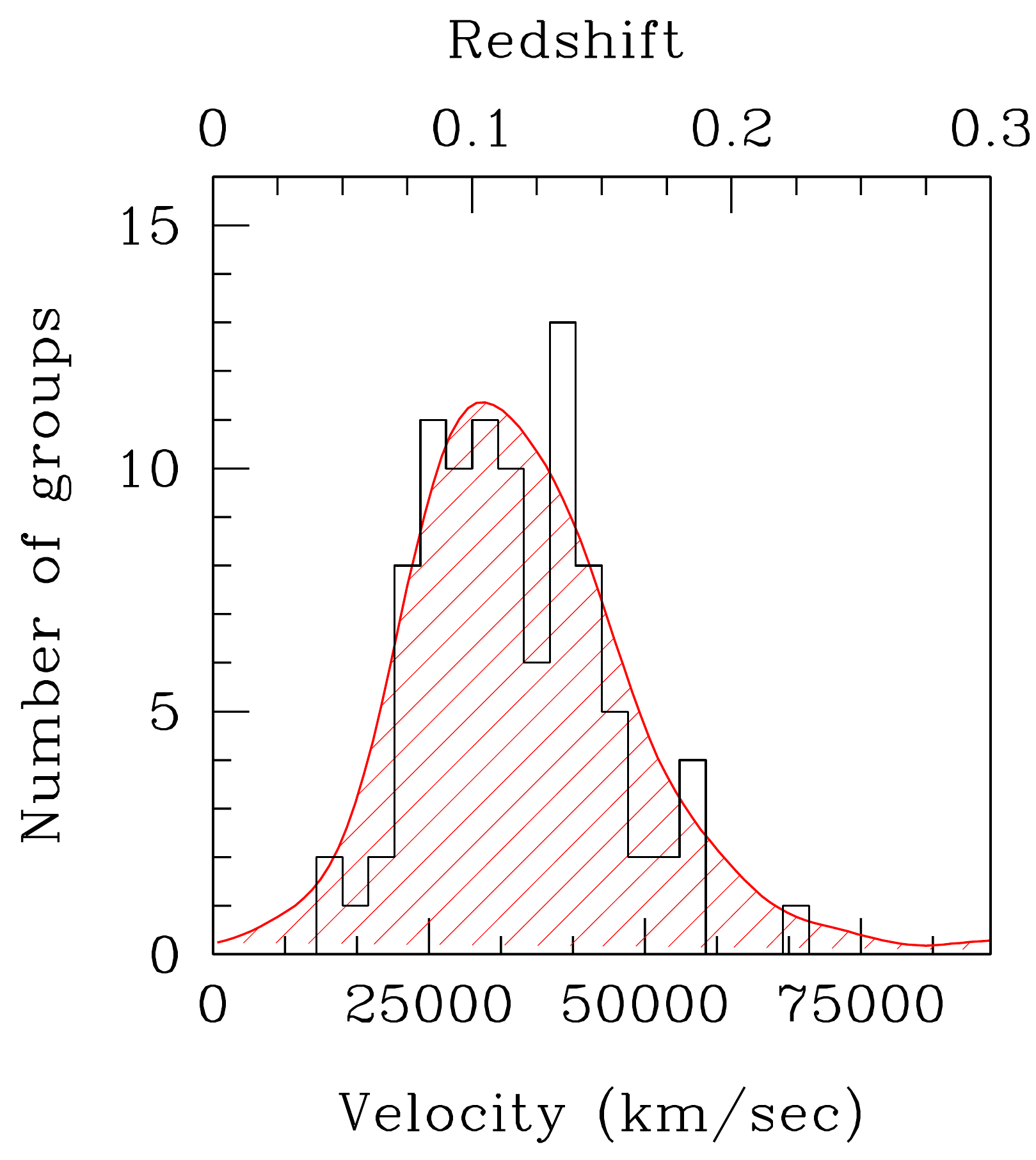}
\caption{Velocity distribution of our confirmed compact groups.}
\label{fig:DPOSS_velo}
\end{figure}

This resulted in 62 isolated compact groups and 34 candidate compact
groups in the vicinity of a cluster or identified with the cluster
core itself. 
The isolated spectroscopically
confirmed groups were labelled {\it class A}, while the others were
put in {\it class B}. To {\it class C} belong all candidate CGs with fewer than three concordant members; 
we note that the majority of {\it class C} groups are composed by couples of
pairs.

\subsection{The small-scale environment}

After distinguishing isolated CGs from those close on the sky to larger-scale
structure, we proceeded to a closer examination of the environment
surrounding our isolated groups.
Using again the last release of the SDSS (DR7), and other literature
sources, as available from NED, we searched for nearby
galaxies within a radius equal to three times the average group radius,
R$_{G}$, (see Sect. 3.6), then a radius equal to 250 kpc, i.e five times the average group
radius and finally a radius equal to 500 kpc, i.e. ten times the average
group radius. 
To ensure that we did not assign random field galaxies to groups,
we also checked the radial velocity distribution of the original group
members and the newly detected galaxies. A new object was
considered an additional member only if its velocity difference from
the median velocity of the group is smaller than 1000\kms.\newline

The goal of this exercise was to investigate 
how many of our isolated CGs are
part of a wider galaxy distribution and how many are real isolated CGs with a high density on the sky. 
We must point out that only part of our sample (see Fig. 3) falls within the area covered by the SDSS
spectroscopic survey, and that the spectroscopic limit of the SDSS is
reached at a magnitude
r$_{petro}$ is 17.7. This means that we do not include in our search all those galaxies whose magnitude is fainter than the
spectroscopic
limit of SDSS, including potential additional group members whose
magnitude ranges from r=17.8 to r=19.0, which is, by construction, the
fainter magnitude limit of our search for group members.  In addition to this, three of our confirmed groups are
completely outside the surveyed areas, while another 13 objects are
located at the edge of the area covered by the survey. 

Despite these limitations, we were able to place constraints on the
local environment of almost all our confirmed groups, using also other sources
of information available from the literature. \newline

\setcounter{figure}{2}
\begin{figure}
\includegraphics[scale=0.5]{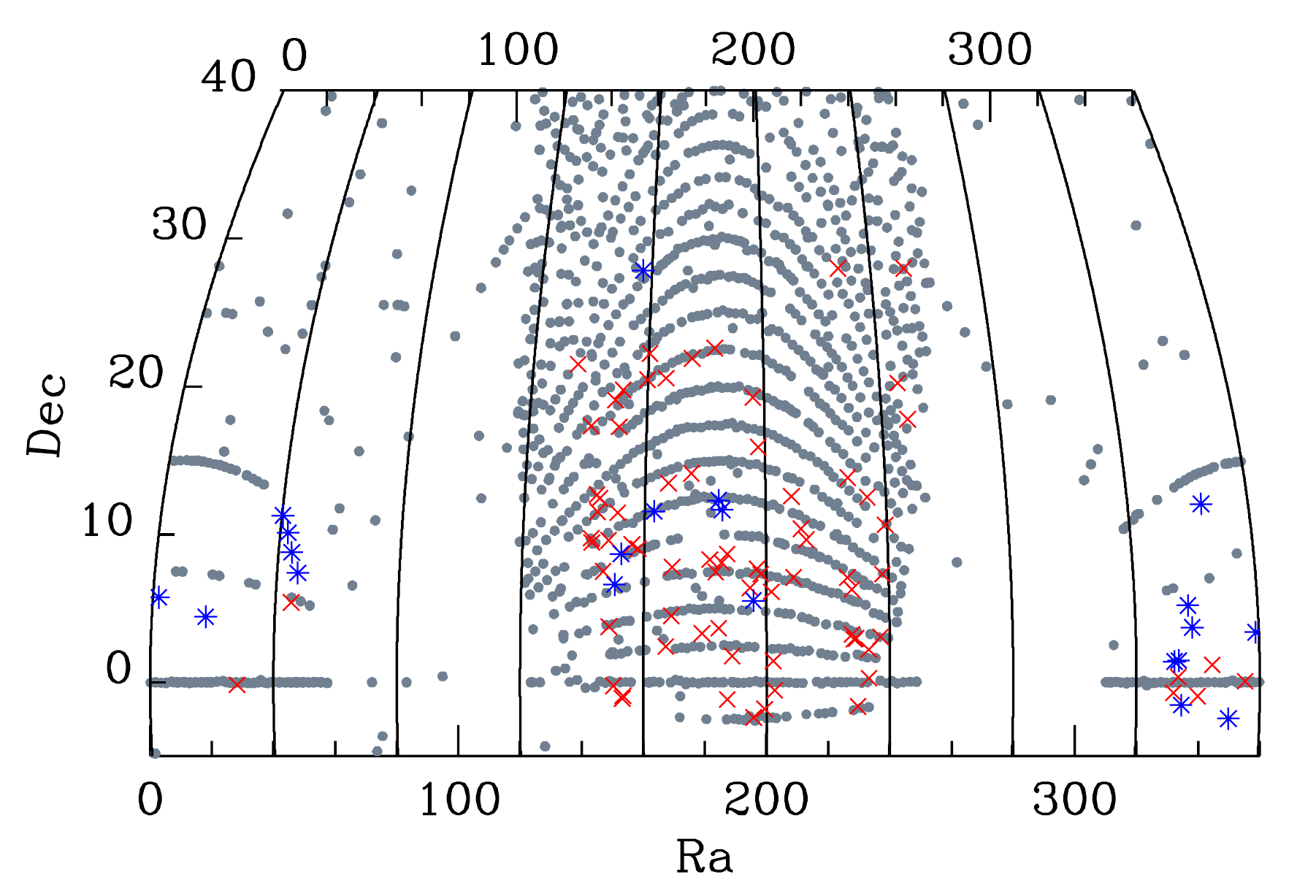}
\caption{Distribution of our confirmed compact groups over the SDSS
  spectroscopic coverage. Crosses represent DPOSS CGs within the SDSS
spectroscopic coverage, while asterisks are objects located at the
edge of a plate or completely outside the surveyed area.}
\label{fig:SDSS_spectro}
\end{figure}

The results of this search can be resumed as follows:
\begin{itemize}
\item{} Six groups belonging to {\it B class} are located at the edge
of a larger-scale structure. From our data, it is impossible to
determine whether these groups are interacting with the cluster, but it would be worthwhile
to perform spectroscopic follow-up observations.
\item{} The remaining  {\it B class}  groups, with the exception of two groups, are either within an
  Abell cluster or coincide with one of the massive clusters identified by
  Koester (Koester et al, 2007). The last two objects, which are
  outside the SDSS area coverage on the sky, are associated
  with an X-ray cluster.
\item{} Groups of class A are a mixed bag of objects: among the total of 62
  spectroscopically confirmed groups, we found 12 with
unusually large radial velocity dispersions, larger than 400\kms, which
is more typical of large groups or poor clusters.
We discuss these 12 objects in more detail below.
\end{itemize}

PCG093220+171954 is composed of
four galaxies. Looking at the velocity distribution, this
group is clearly composed of two pairs of galaxies at similar redshift, surrounded by other galaxies. 
PCG094321+122625 has a chain-like geometry, i.e. all member galaxies 
are aligned with each other on the sky. However the redshift
distribution
of its member galaxies is clearly binomial, revealing this group as a
couple
of pairs close to each other on the sky.
A similar situation, albeit with a different geometry, happens for
PCG095527+034508 and PCG153046+123131, i.e. all of them are
composed of two close pairs at similar redshift. These
two pairs may merge with each other in the future, forming a new group of
galaxies, but here these groups were discarded from the final sample.
PCG100644+112806  and PCG155341+103913 appear instead to consist of
a pair of galaxies with an infalling galaxy, which is quite
distant from the other two in velocity space, albeit still within the
1000\kms limit.
These objects, which passed the first selection thanks to the small
velocity difference between the two pairs, or between the couple and
the third galaxy, were excluded from the sample.\newline

PCG121738+121833 and PCG151057+031443
are part of a larger structure on the sky: other galaxies, of magnitude comparable
to that of the brighest group galaxy, surround the original members.
These are within wider structures of the order of 1 Mpc wide. We define
this type of group a {\it loose group}.\newline

In addition PCG161009+201350, PCG221442+012823 and PCG225807+011101  
are surrounded by other nearby galaxies,
but the original members are closer to each other on the sky than the surrounding
members. This kind of association is often classified as a loose group, but we call it a
{\it core+halo} group, following Ribeiro et al., 1998.\newline

It is likely that our selection algorithm selected this kind of
structure because the original four members were somehow closer 
together on the sky than the other galaxies belonging to the
structure, triggering a positive detection. 

No companions can be found for PCG222633+051207, despite many galaxies
being identified in the field, hence it retains its original
classification of compact group. However, many bright galaxies are detected in the
acquisition images, implying that the environment surrounding this
group is a
markedly rich one. We then decided to flag this group as suspect
and more likely to be a loose group.

Another candidate compact group that was not observed by us, but
identified using SDSS data, PCG130257+053112, also contains other four
galaxies within 500kpc and $\Delta \le$ 1000\kms, and was classified as a {\it core+halo} group, despite its
small velocity dispersion.

This means that from the original 62 isolated and spectroscopically
confirmed compact groups, we lose 6, because either close couples or pairs with
a third close member, and another 7 are larger-scale structures,
i.e. loose or core+halo groups, leaving us with 49 {\it class A}, {\it bona fide}
compact groups.

\subsection{Group density on the sky}

To ensure that our confirmed groups are really dense concentration of galaxies on the sky, 
we apply a density criterion, according to the formula (Ribeiro et al, 1998)

\begin{equation}
\mathrm{\rho}  = \frac{3\mathrm{N}}{4\pi\mathrm{R^{3}}},
\end{equation}

where N is the number of galaxies in the group that are spectroscopically confirmed and R is 
the group radius in Mpc. A compact group is considered such when its galaxy density is equal 
to or above 10$^3$ galaxies Mpc$^{-3}$.
The total distribution of the measured group densities is shown in Figure 4, which compares the 
full confirmed
group sample with the isolated one and shows the cut-off line at 10$^3$ galaxies Mpc$^{-3}$.

\setcounter{figure}{3}
\begin{figure}
\includegraphics[scale=0.35, angle=-90]{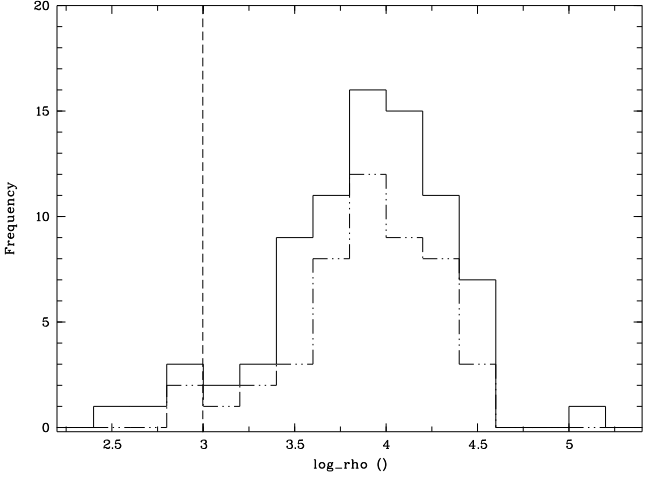}
\caption{Distribution of the group galaxy density for the whole sample (continuous line)
and the {\it class A}
sample (dotted-dash line). The vertical line is the cut-off density limit at 10$^3$ galaxies Mpc$^-3$. The
average density is 1.2 x 10$^4$ galaxies Mpc$^{-3}$.}
\label{fig:density}
\end{figure}

The density criterion is failed  by two {\it class A} CGs,
PCG093310+092639 and 
PCG130926+155358, both have small velocity dispersions and very 
sparse configurations on the sky: these have been discarded from
the final sample. This leaves us with only 47 CGs in {\it class A}.\newline

If we examine the {\it class B} groups, we find that three of them, PCG101113+084127, PCG121346+072712, 
and PCG151329+025509 have a density below the limit set for CGs: all are in the middle of
larger structures, hence we may question how this density would change if the whole larger-scale structure were taken
into account.\newline 
A special case however is PCG151329+025509, which is in a very
perturbed area with another three nearby interacting 
galaxies: it is at the same redshift as the cluster
MaxBCG J228.37839+02.91616 (Koester et al., 2007), and it is unclear
whether this is a distant cluster with substructures.

\subsection{The final sample of compact groups}

Our final candidate CGs classification is then as follows: all the compact group candidates that are not close
on the sky to a larger-scale structucture, and fulfil the density criterion are classified as {\it class A}.
Within this class, three categories of objects are identified:
\begin{itemize}
\item{a} Real compact groups: 47 final confirmed targets
\item{b} Loose groups: 3 objects
\item{c} Core+halo groups: 4 objects
\item{d} Close couple of pairs or a close pair with a third galaxy
  within the $\Delta$v $<$ 1000\kms limit: 6 objects
\end{itemize}

As already mentioned in Sect. 3.2, objects belonging to the last
category were removed from the final list.
We also excluded {\it loose}, {\it core+halo} groups, and those groups
failing the density criterion from the
final list, considering as {\it bona fide} compact groups only those
at point {\it a }.\newline

Candidate compact groups that are close on the sky to a large-scale structure and fulfill the density
criterion are classified {\it class B} and represent 25$\%$ of the
whole sample. Within this class, two subcategories are identified:
\begin{itemize}
\item{} Groups that are affected by the larger-scale structure, as
  traced by an larger than normal velocity dispersion, higher virial
  mass.
\item{} Groups that, despite their closeness to a larger structure,
  seem to retain their identity and have velocity dispersion, radius,
  and mass typical of compact groups. It is likely that these groups will
  interact with the other structures in the distant future, but at the moment they can be
considered as independent structures.
\end{itemize}

If we consider only {\it class A} objects as real compact groups, the
success rate of this survey is 34$\%$.
The whole sample of observed and confirmed compact groups, with the
classification for each group is listed in Table 1.

\subsection{Internal dynamics and mass estimates}
To understand how our distant CGs, be them isolated or closer
to a large-scale structure, compare with nearby ones, we proceed to 
measure the characteristic properties, i.e. the three-dimensional (3D)
velocity dispersion, crossing
time, and mass, mass-to-light ratio, using the luminosities derived as
explained in Sect. 2.3.

For the 3D velocity dispersion, we use the same
equation used in Hickson et al. (1992), the crossing time being defined as:

\begin{equation}
t_\mathrm{c} =  \frac{4}{\pi} 
                \frac{R}{\sigma_{3D}}
\end{equation}

where R is the median of galaxy-galaxy separations and $\sigma_{3D}$ is 
the 3D velocity dispersion. The dimensionless crossing time,
shown in column 5 of Table 2, 
$H_\mathrm{o}t_\mathrm{c}$, spans from 
0.002 to 0.135, with a median value of 0.020, which is slightly larger than the value measured for HCGs, 0.016.

The median observed velocity dispersion is 273 \kms, while
the 3D velocity dispersion is 382 \kms, substantially larger than
that found for the HCGs (Hickson et al., 92).

If we restrict ourselves only to isolated groups, i.e. the {\it class A} ones,
the median crossing time becomes
$H_\mathrm{o}t_\mathrm{c}$ = 0.024,
50$\%$ longer than the crossing time measured for HCGs but still half
of the median value measured for SCGs, 0.051.
The median radial velocity dispersion is $\sigma_{r}$ = 188\kms,
while the 3D velocity
dispersion is $\sigma_{3D}$ = 250\kms, very similar to that
found for nearby CGs.\newline
If we consider the groups associated to larger structures, excluding those
in the middle of a cluster, we measure a median radial
velocity dispersion of $\sigma_{r}$ = 310\kms, while the 3D velocity
dispersion is $\sigma_{3D}$ = 433\kms, which is about 1.6 times larger than
the value measured for the isolated groups, in agreement with the result of Einasto et
al. for loose groups closer to large-scale structures on the sky.

A comparison of the crossing time and velocity dispersion of the whole
DPOSS sample and the isolated DPOSS compact groups is shown in Figure
5. A {\it k-s} test shows that the two populations are different at
a confidence level of 97$\%$.

\setcounter{figure}{4}
\begin{figure}
\includegraphics[scale=0.4]{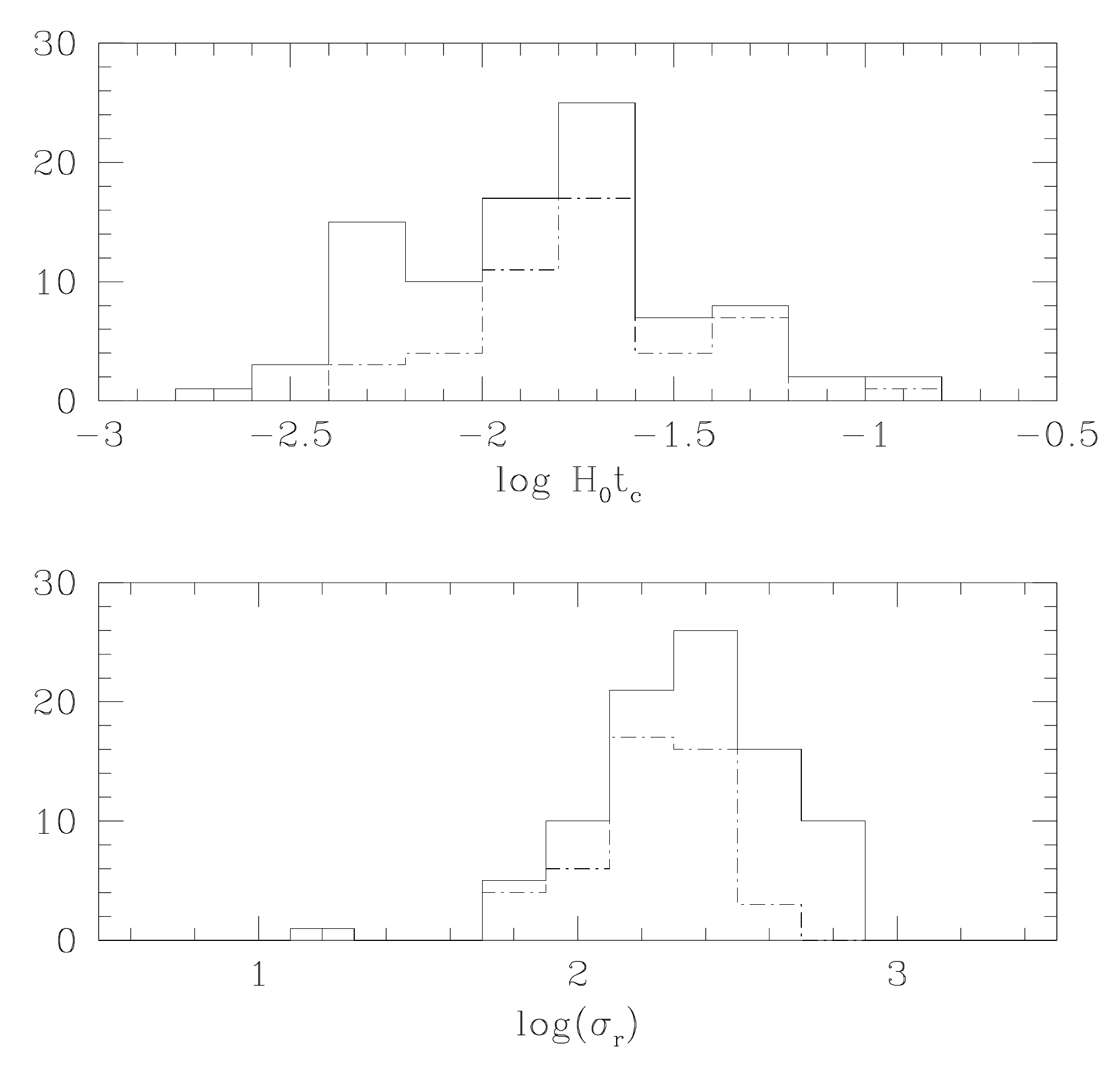}
\caption{Top panel: distribution of the crossing time for the whole
  DPOSS sample (continuous line) versus the {\it class A} groups
  (dot-dashed line). Bottom panel: Same distribution for the radial velocity dispersion.}
\label{fig:dposs_vs_dpossA}
\end{figure}

\vspace{0.5cm}

For the mass estimate, we use different estimators, the virial and the
projected mass. 
The expression for the virial mass is given in Equation 4, which
is valid only under the assumption of spherical symmetry.

\begin{equation}
\mathrm{M_{V}}  = \frac{3\pi\mathrm{N}}{2\mathrm{G}}
                  \frac{\mathrm{\Sigma_{i}V_{zi}^2}}{\mathrm{\Sigma_{i<j}1/R_{ij}}}
\end{equation}

where $R_{ij}$ is the projected separation between galaxies i and j,
here assumed to be the median length of the 2D
galaxy-galaxy separation vector, corrected for cosmological effects; N is
the number of concordant galaxies in the system, and $V_{zi}^2$ is the velocity component
along the line of sight of the galaxy {\it i} with respect to the centre
of mass of the group.
As observed by Heisler et al. (1985) and Perea et al. (1990), the
use of the virial theorem produces the best mass estimates, provided
that there are no interlopers or projection effects.  In case one of
these two effects is present,  the current values should be considered as
upper limits to the real mass.

Another good mass estimate is given by the projected mass estimator, which is defined as

\begin{equation}
\mathrm{M_{P}} = \frac{f_{P}}{GN} 
                 \mathrm{\Sigma_{i}V_{zi}^2}\mathrm{R_{i}}
\end{equation}

where $R_{i}$ is the projected separation from the centroid of the system,
and f$_{P}$ is a numerical factor depending on the distribution of the orbits
around the centre of mass of the system.

Assuming a spherically symmetric system for which the Jeans hydrostatic equilibrium
applies, we can express {\it f$_{P}$} in  an explicit form (Perea et al., 1990). 
Since we lack information about the orbit eccentricities, we estimate the mass
for radial, circular, and isotropic orbits and the corresponding expressions for
M$_{P}$ are given in Equations 6, 7, and 8 respectively as

\begin{equation}
\mathrm{M_{P}} = \frac{64}{\pi G}
                 {<V_{z}^2 R>}
\end{equation}

\begin{equation}
\mathrm{M_{P}} = \frac{64}{3\pi G}
                 {<V_{z}^2 R>}
\end{equation}

\begin{equation}
\mathrm{M_{P}} = \frac{64}{2\pi G}
                 {<V_{z}^2 R>}
\end{equation}
 
where R is the median length of the 2D
galaxy-galaxy separation vector.\newline

The results for the four estimators all agree
quite well with each other, and the reported value for the mass in
column 6 of Table 2 is the average of all four estimates.
The averaged values have been used for the estimate of the M/L ratio
in column 8 of Table 2. We note that revised values for some
groups from
paper I are published here, to correct a former error in
velocity measurements owing to problematic wavelength calibrations.

The group masses vary from $2.67\times 10^{10}$ to $1.34\times 10^{14}$ $M_\mathrm{\sun}$, 
with an average value of $\sim$ $1.67\times 10^{13}$ $M_\mathrm{\sun}$.
The M/L ratio varies from 0.38 to 2476, with an average value of 192,
which is larger than that reported for
HCGs, and similar to that measured for loose groups of galaxies.

If we restrict ourselves to the isolated compact groups, i.e. the {\it class A} ones, the average values
of mass and mass-to-light ratio are both lower, i.e. M = $6.6\times 10^{12}$ $M_\mathrm{\sun}$, 
and $M/L_{B}$ = 80, which are very similar to the values measured for compact
groups in the nearby universe.
Groups close to larger-scale structures have an average mass of M =
$2.3\times 10^{13}$ $M_\mathrm{\sun}$,
and $M/L_{B}$ = 262. The measured mass is $\sim$ 2.5 times the value
measured for isolated groups, once more in agreement with the result
found by Einasto et al. (2003).

\subsection{Radius distribution}

The typical radius measured for compact groups in the nearby universe is of the order of 50 kpc, or less.
We measured the group radius, R$_{G}$, as the median of the galaxy-galaxy
separation within the spectroscopically confirmed members of each group.
In Figure 6, we show the histogram of the radius distribution for our whole sample. The peak of the
distribution is at 50 kpc, which is quite consistent with the value measured for
compact groups. The mean value does
not change if only isolated groups are considered. This is expected, because this
radius is measured taking into account only the original,
spectroscopically confirmed, group members.

\setcounter{figure}{5}
\begin{figure}
\includegraphics[scale=0.3, angle=-90]{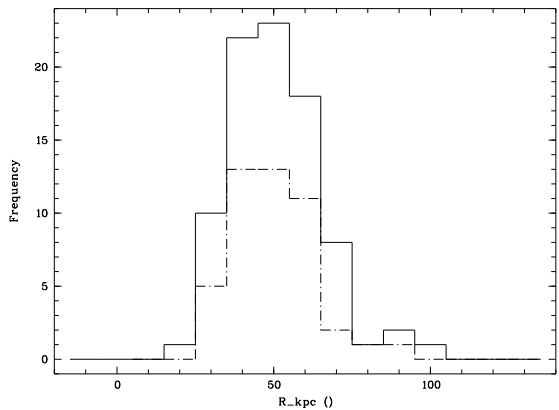}
\caption{Distribution of the median group radius for the whole DPOSS sample of confirmed compact groups
(continous line) and the isolated one (dot-dashed line).}
\label{fig:radius}
\end{figure}

However, if one computes the virial radius for our objects, as

\begin{equation}
R_\mathrm{vir} =  \left(\frac{3M}{4\pi\rho_{200}}\right)^{1/3} 
\end{equation}

where $\rho_{200}$ is 200 times the critical density of the Universe
and M is the group virial mass,
the average value differs for the whole sample of our confirmed groups
and the isolated ones, being R$_{200}$=468 kpc and R$_{200}$=310 kpc, respectively.

\subsection{The population of galaxies in the groups}

We studied the galaxy morphological types for each group, defining a
galaxy to be 
late type if its H$\alpha$ equivalent width
is larger than 6$\AA$, following Ribeiro et al., 1998. This choice is
motivated by the shallowness of our acquisition images and their
varying depth, which do not allow a reliable photometric analysis.
This criterion probably allows active galaxies to contaminate the
sample; however, the measured fraction of strong
AGN (Sy galaxies) in CGs is of the order of 12$\%$ (Martinez
et al., 2008), and only one galaxy in 238 observed is a broad-line active galaxy.
Of 370 galaxies within our group population, we find that 72
are spiral galaxies, i.e. 19$\%$ of the total. 
However, if we subdivide our sample into groups close to a large-scale
structure, {\it class B}, and isolated groups, {\it class A}, the
percentage of spiral galaxies changes to 14$\%$ and 24$\%$,
respectively, i.e. isolated compact groups contain a larger fraction
of late-type galaxies.\newline
The spiral fraction is a function of the crossing time as
shown in Fig. 7. 

\setcounter{figure}{6}
\begin{figure}
\includegraphics[scale=0.5]{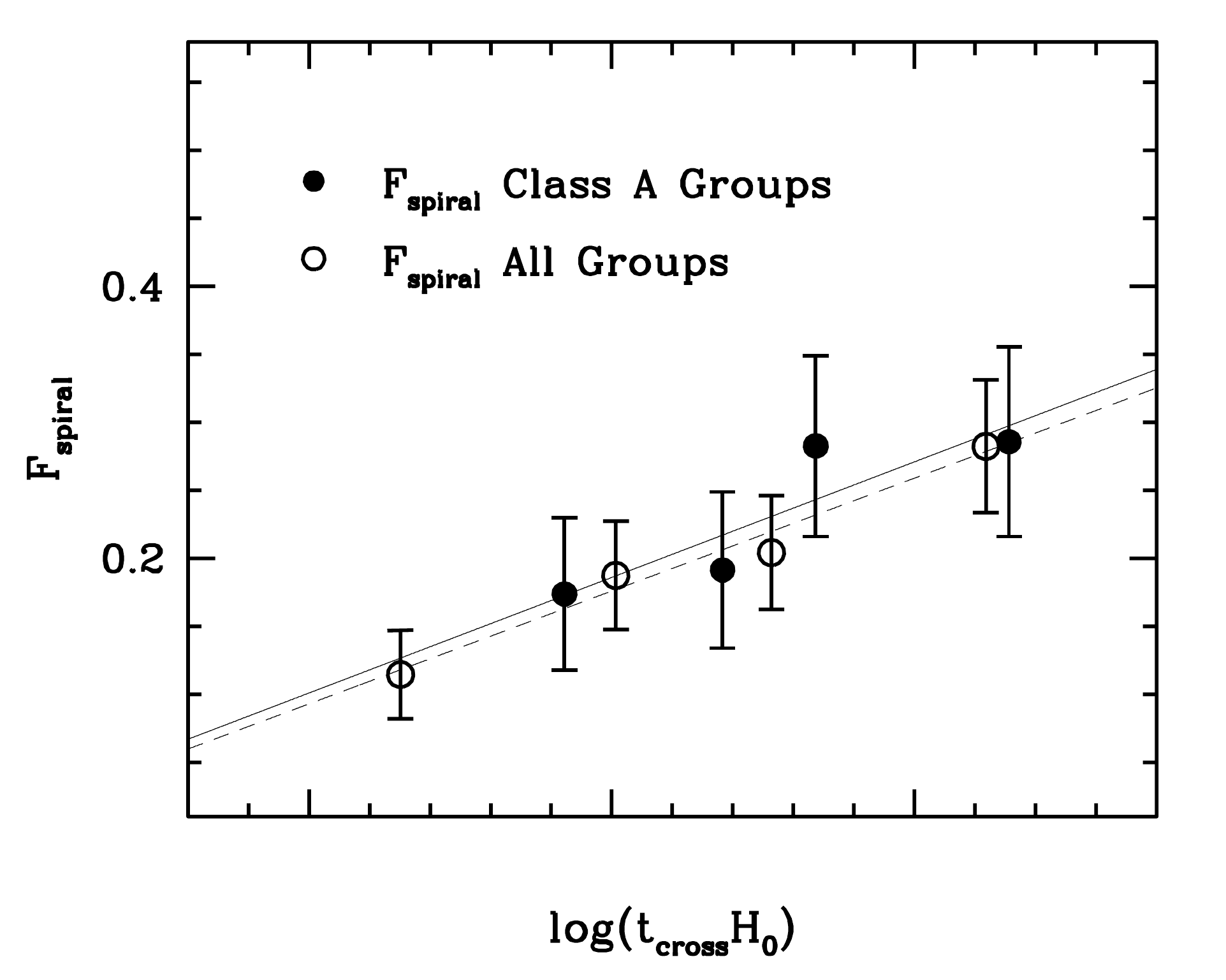}
\caption{Spectroscopically selected fraction of spiral galaxies as a function of the crossing
  time for all spectroscopically confirmed groups and for
the {\it class A} groups only. The solid and dashed lines represent
the fits for the {\it class A} groups and the whole sample, respectively.}
\label{fig:spirals}
\end{figure}

The spiral fraction increase can be fit by
a linear slope

\begin{equation}
\mathrm{f_s}  = 0.508+0.166(\pm0.019)\times log(H_\mathrm{o}t_\mathrm{c})
\end{equation}

This trend does not change if only isolated groups are considered, becoming

\begin{equation}
\mathrm{f_s}  = 0.526+0.170(\pm0.064)\times log(H_\mathrm{o}t_\mathrm{c})
\end{equation}\

The spectroscopic fraction of spiral galaxies is the smallest measured
so far in compact groups of galaxies; the number is well
below the fraction measured for HCGs, 49$\%$, and for SCGs,
69$\%$ (Pompei et al., 2003). Such large difference might be partly
caused by our use of a spectroscopic morphological criterion, while the quoted spiral
fraction for nearby compact groups was derived from deep photometric
studies.
However, if only a spectroscopic criterion is used for HCGs, the
spiral fraction remains quite high, of the order of 40$\%$ (see Fig. 3 of
Ribeiro et al., 1998). Hence, it seems that our confirmed compact groups
have indeed a smaller fraction of late-type galaxies than HCGs.

\subsection{Comparison with other surveys}

We checked the literature to compare our results with other catalogues of groups of galaxies: since
the release of the SDSS DR7, two major catalogues of galaxy groups have been published, one by McConnachie et al.
(2009) and another by Tago et al (2010).
We compared our whole observed sample of 138 groups with catalogues A and B from McConnachie and Table 2 from Tago et
al. The geometrical center of each group was used in all the catalogues and a search radius of 30$\arcsec$ on the sky
was used, returning a total of 23 matches from McConnachie and 5 from
Tago et al.  The choice of the radius was a compromise between the two
different definitions of group center of ourselves and McConnachie, as we
both used the geometrical radius, and Tago et al., who used a
different method. When redshift measurements were available,
a good agreement was found in all cases, except five (see Table 3). In these cases, no more than two redshifts were available from SDSS
data, while our own observations proved that the candidate compact
group consisted of a couple of pairs at two different
redshifts. 

\setcounter{table}{2}
\begin{table*}
\caption{\label{group_check} Associations between the group catalogue of
  McConnachie et al. and our confirmed group sample. The
  second column lists the average spectroscopic redshift measured
  by us: no redshift means that the group was not confirmed
  spectroscopically.
 The fourth and fifth colums list the available
  photometric and spectroscopic redshift from the SDSS data as quoted
  by McConnachie et al., 2009. In the final column, the separation
  between the objects in arcsec is given.}
\begin{tabular}{l l l l l l }\hline\hline
DPOSS group name   &  z$_{DPOSS}$ & SDSS group   & z$_{phot}$  &
z$_{spectro}$& Separation \\ \hline
PCG015254-001033 &	0.081 &	SDSSCGA00210 &	0.55	& 0.081 & 4.13\\
PCG091524+213038 &	0.134 &	SDSSCGA00165 &	0.38	& 0.134 & 6.37\\
PCG092231+151104 &	   -	  &    SDSSCGA00472 &	0.45	& 99.999 & 7.89\\
PCG095507+093520 &	0.144 & 	SDSSCGA04 &	0.31	& 99.999 & 1.98\\
PCG100102-001342 &	0.092 &	SDSSCGA00126 &	0.42	& 0.092 & 2.71\\
PCG104530+202701 &	0.130 &	SDSSCGA04 &	0.35	& 99.999 & 5.64\\
PCG104841+221312 &	0.045 &	SDSSCGA00294 &	0.72	& 99.999 & 10.39\\
PCG112051+074439 &	   -	  &	SDSSCGA04 &	0.65	& 99.999 & 9.70 \\
PCG114233+140738 &	0.125 &	SDSSCGA00065 &	0.35	& 0.1 & 5.72\\
PCG114333+215356 &	0.132 &	SDSSCGA00075 &	0.31	& 0.132 & 13.75\\
PCG115610+031802 &	0.070 &	SDSSCGA00048 &	0.27	& 0.072 & 5.32 \\
PCG121516+153400 &	   -     &	SDSSCGA00332 &	0.53	& 99.999 & 9.08 \\
PCG125835+062246 &	0.082 &	SDSSCGA00086 &	0.37	& 99.999 & 3.68 \\
PCG131211+071828 &	0.093 &	SDSSCGA00034 &	0.27	& 99.999 & 1.11 \\
PCG132826+012636 &	0.078 &	SDSSCGA00280 &	0.6	& 0.079 & 4.23 \\
PCG140026+053457 &	   -     &	SDSSCGA00098 &	0.33	& 0.035 & 0.75 \\
PCG150457+070527 &	0.092 &	SDSSCGA00300 &	0.56	& 0.092 & 4.75 \\
PCG151329+025509 &	0.134 &	SDSSCGA00302 &	0.57	& 0.135 & 8.42 \\
PCG151833-013726 &	0.063 &	SDSSCGA00474 &	0.68	& 99.999 & 6.72 \\
PCG154930+275637 &	    -    &	SDSSCGA01258 &	0.79	& 99.999 & 18.90 \\
PCG155024+071836 &	0.101 &   SDSSCGA00393 &	0.54	& 0.102 & 6.77 \\
PCG162259+174703 &	0.113 &	SDSSCGA00446 &	0.73	& 99.999 & 10.85 \\
PCG220748-004159 &	0.109 &	SDSSCGA00475 &	0.59	& 0.11 &	7.51 \\
\hline\hline
\end{tabular}
\end{table*}

The DPOSS cluster catalogue (Lopes et al., 2004) was also searched for
possible associations between our candidate CGs and 
larger-scale structures from the same survey. Here the results were somewhat mixed, because of the disagreement
between the quoted photometric redshifts from the DPOSS and our spectroscopic redshifts. Owing to the robustness
of the spectroscopic measurements, in particular at such low redshifts, we conclude that our redshift estimates are more
reliable and reassign the same distance to the DPOSS cluster that is close on
the sky to
our candidate DPOSS compact groups and the group itself (see Table 1
for confirmed associations).

\section{Discussion}

From our analysis, it becomes clear that our original sample of
candidate CGs consists of a mixed bag of objects,
a significant part of which is embedded in a large-scale structure.
Most of the rejected candidates are formed by two pairs of galaxies
close together on the sky.
Hence, a first result is that finding compact groups at intermediate redshift is not a trivial
business, even when using search criteria based on the original
Hickson's ones and modified to take into account the larger
distance. Once the candidate groups have been identified, it is of crucial importance not only to confirm
spectroscopically all the candidate member galaxies, but also to study
in detail the surrounding environment of the confirmed groups, to have
a good understanding of what kind of objects are being
observed. This should return a reasonably clean sample of isolated compact groups.
Even with these precautions, our final sample represents an upper limit to the amount of
isolated compact groups, partially because of the incomplete coverage
of the SDSS in the areas observed by ourselves and partially because of the
limited spectroscopic coverage for objects fainter than r$_{petro}$
=17.7.
Despite all these limitations, we have found a fraction of confirmed isolated
groups of 34$\%$  of the total number of groups originally
observed, which is larger than the fraction measured for a subsample of 
HCGs (de Carvalho et al., 1994, 23$\%$). We wonder, however, what the
real fraction for HCGs would have been if the whole 92 confirmed groups
catalogue had been studied by de Carvalho et al. 1994\newline

Our confirmed isolated CGs have a median crossing time equal to
about one tenth
of the light travel time to us. Assuming the crossing time as an
estimate of the group age, we conclude that we do not observe at
higher redshift the same
population of groups as we observe in the nearby universe; that is 
we must be looking at a different population of
compact groups. If we were to assume that all these groups will continue to
exist in isolation without further perturbation from the external
environment, the most likely final product of such groups would be an
isolated early-type galaxy. According
to the model of Barnes (1989), isolated CGs are likely to
evolve into a single isolated elliptical galaxies in a few crossing
times, hence we expect these z=0.1 objects to be the progenitors of
present-day ellpticals.\newline
To determine whether this conclusion is realistic, at least
in qualitative terms, we compared the volume density of our confirmed
CGs to the volume density of isolated early-type galaxies
drawn from several literature works in the nearby universe. 
We assumed that the density
distribution of our compact groups in space is uniform across the surveyed
area, and that our sample is complete all the way to our average
redshift; we were then able to estimate an average volume density of CGs
based on Eq. 1 of Lee et al., 2004. Our isolated compact
groups have a density of 1 x 10$^{-6}$Mpc$^{-3}$, similar to the values
measured for clusters of galaxies (Bramel et al, 2000) and fossil
groups (Santos et al., 2007; Jones et
al., 2003). \newline
Assuming that the percentage of early-type galaxies is $\sim$
18$\%$
of all field galaxies (starting from an estimated 82$\%$ fraction of
spirals in the field, as quoted by Nilson et al., 1973 and Gisler
et al., 1980), we applied the same Eq. 1 to the samples of Allam et
al, 2005; Giuricin et al. 2000 and AMIGA (Verley et al., 2007).
Assuming as an average redshift for each survey the one quoted in each
paper, we found that the volume density of early-type galaxies is of
the order of a few $\times$ 10$^{-5}$Mpc$^{-3}$. If we adopted the
maximum redshift of each survey for which completeness is
claimed, the number goes down to a few $\times$ 10$^{-6}$Mpc$^{-3}$. \newline
We conclude from these numbers that it is perfectly plausible that
all our isolated compact groups will end up as early-type galaxies in
the field with no overpopulation problem.\newline
Other indirect evidence can further strenghten our
reasoning: likely candidates for such end-product from groups
were previously observed by the MUSYC-YALE survey (van Dokkum et
al., 2005). They are isolated {\it red and dead} galaxies, which, on deeper
inspection, contain extended tidal tails and shells composed mainly of
stars, with a small amount of residual gas, the so-called {\it dry mergers}.\newline
If we examine exisisting spectroscopic studies on galaxies in CGs,
we find that observations of nearby isolated early-type galaxies (Collobert et al.,
2006) have shown that the most massive galaxies in low density
environments
have abundance ratios similar to those of cluster galaxies. Mendes de Oliveira
et al., 2005 also demonstrated that early type galaxies in HCGs are
generally old. On the other hand,
early-type galaxies in the field with small central velocity
dispersions have properties that are consistent with extended episodes of star
formation (Collobert et al., 2006), as if coming from a past of
multiple interaction and slow buildup.\newline
Available X-ray observations of CGs reveal a wide range of the
X-ray diffuse emission, with a slight tendency for spiral-rich groups
to have a small amount of X-ray emission (Ponman et al., 1996). 
An opposite trend is tentatively detected for the HI content
(Verdes-Montenegro
et al., 2001; Pompei et al., 2007).\newline
On the basis of these observations, the following scenario can be envisioned:
galaxies belonging to more massive CGs evolve mainly within
the group environment, giving rise to either a fossil group, i.e. a
massive elliptical galaxy, surrounded by several dwarf galaxies, or to
a field elliptical, whose abundance ratios are similar to those observed in
clusters. As the galaxies have already evolved within the group, no or
very little young stellar population should be present and they should
continue to evolve via {\it dry mergers}.
In both cases, the aforementioned X-ray emission is expected.\newline
Galaxies belonging to less massive compact groups evolve in a more gradual
way, probably by means of stripping of gas from each other through harassment,
giving rise to several episodes of nuclear star formation, and diffuse
HI emission within the group potential. 
They will likely end up as a single isolated
early-type galaxy with a younger stellar population than those
observed in clusters of galaxies and possibly extended tidal features
composed of a small percentage of gas and a high percentage of stars,
which are the mute
witnesses
of the final merger of the compact group in a single galaxy. No or
very little X-ray emission is expected in this case, because almost all
the existing gas would have been either exhausted in the former
episodes of star formation or lost into the intergalactic medium.
The dominant factors in shaping the different evolutionary pattern
seem to be the velocity dispersion of the group and  its initial
mass.\newline

Moving toward the {\it class B} groups, we find that among 34, nine groups are
close to larger-scale structures but not embedded within them; four of these
groups have characteristics that are very similar to isolated compact
groups, while the others are likely to be affected by the cluster potential,
as deduced from their larger velocity dispersion and group
position with respect to the cluster. \newline
The percentage of DPOSS groups closer on the sky to larger-scale
structures is 25$\%$ (34 over 96 confirmed groups), in agreement with
what has been found by Andernach \& Coziol, 2005.
The higher mass and larger velocity dispersion of groups in the
proximity of larger scale structures support the findings of Einasto
et al., of the hierarchical formation of galaxies. In this scenario, it
can be postulated than less massive groups formed in lower density
regions of the cosmic filaments.


\section{Conclusions}

We have presented our results of a spectroscopic survey
of compact groups at a median redshift of z$\sim$=0.12, i.e. a factor
of ten larger than any previous study.
These can be summerized as follows:
\begin{itemize}
\item{} Among a total of 138 observed groups, we confirm 96 compact
  groups with 3 or more accordant members.
\item{} Forty-seven of the confirmed groups are isolated groups on the sky,
  i.e. a success rate of 34$\%$.
\item{} The average mass, mass-to-light ratio, crossing time, radius,
  and velocity
  dispersion of our isolated compact groups are very similar to the
  values obtained for compact groups in the nearby Universe.
These values are different from those measured for groups close to a
larger-scale structure on the sky.
\item{} Isolated compact groups tend to have a longer crossing time
  and a higher fraction of spiral galaxies.
\item{}  The volume density of isolated compact groups is consistent with the hypothesis that all of
  them will conclude their life as a single isolated early-type
  galaxy. Depending on the original mass and velocity dispersion of
  the group, we expect the final merger product to resemble a cluster
  or a field galaxy, with or without an extended X-ray
  halo. 
\item{} Nine confirmed groups are larger-scale structures, loose groups,
  or {\it core+halo} groups, and will likely behave differently from
  an isolated compact group.
\item{} Six objects were discarded, because they were close couples of
  pairs in redshift space and during the first selection were
  mistaken for compact groups. It is possible that such close couples
  of pairs can come together to form a group, but this is at the moment a
  matter of speculation.
\item{} Thirty-four of the confirmed groups are close on the sky to a
  larger-scale structure, 
to which they might be associated. Of these groups,
  four are still retaining their identity, while five others are probably
  already being perturbed by the cluster potential.
The percentage of association between groups and larger-scale clusters
is in agreement with that found by Andernach \& Coziol in 2005.

\end{itemize}

We stress that each study of compact groups or any specific
environment, needs careful to incorporate consideration of the
surrounding larger-scale, 
in order to have a clear understanding of the kind of sample
one is dealing with. 

\begin{acknowledgements}
Sincere thanks go to the glorious La Silla Science Operations
Team from 2004 to 2008, among them: Ivo Saviane, Gaspare Lo Curto,
Valentin Ivanov, Julia Scharwaecther, Jorge Miranda, Karla Aubel,
Duncan Castex, Manuel Pizarro, Monica Castillo, and Ariel Sanchez without
whom none of these data would have been acquired and without whose
company, patience, and fun none of these data would have gone into publication.\newline
Special thanks also to the NED team, for the impressive database and
for their prompt and professional
answers to email queries.\newline
EP wishes to acknowledge the ESO Director General Discretionary Fund
(DGDF) program for two visits
at Milan Observatory and INAF for an extended visit to Milan Brera
Observatory in 2011, which
allowed the completion of this paper. 
\end{acknowledgements}

\setcounter{table}{0}
\begin{longtable}{c c c c c c c}
\caption{\label{group_prop} List of our observed compact groups,
  group coordinates, number of member galaxies, average redshift, and
  our group classification as described in the text}\\
\hline\hline
Group name                &   RA (2000)   & DEC (2000)     & n & cz (\kms) & classification & Notes \\ \hline
PCG001029+175017  & 00 10 29.85 & +17 50 17.16 & 4 & -  &  C & \\
PCG001108+054449  & 00 11 08.97 & +05 44 49.13 & 4 & 43754$\pm$ 35 & A&\\
PCG011206+042617  & 01 12 06.92 & +04 26 17.20 & 3 & 32921$\pm$89 & A &\\
PCG015254-001033  & 01 52 54.30 & -00 10 33.53 &  4 &24417$\pm$66 & A &\\
PCG025234+111647 & 02 52 34.21 & +11 16 47.43 & 4 & 35038$\pm$92 & A &\\
PCG025903+100636 & 02 59 03.73 & +10 06 36.61 & 3 & 35443$\pm$77 & A &\\
PCG030301+052405 & 03 03 01.05 & +05 24 05.36 & 3 & 30477$\pm$84 & A &\\
PCG030352+084700 & 03 03 52.37 & +08 47 00.60 & 4 & 27066$\pm$76 & A &\\
PCG031139+072404 & 03 11 39.92 & +07 24 04.18 & 4 & 43495$\pm$100 & B
& retains its identity\\
PCG031232+072011 & 03 12 32.00 & +07 20 11.90 & 4 & - & C & \\
PCG091524+213038 & 09 15 24.57 & +21 30 38.81 & 5 & 40093$\pm$88 & B & infalling?\\
PCG092231+151104 & 09 22 31.40 & +15 11 04.24 & 4 & - & C & \\
PCG093220+171954 & 09 32 20.28 & +17 19 54.37 & 3 & 41684$\pm$91 & PI & \\
PCG093226+094339 & 09 32 26.01 & +09 43 39.87 & 4 & 22800$\pm$49 & B &
Abell 819\\
PCG093310+092639 & 09 33 10.28 & +09 26 39.33 & 3 & 41309$\pm$93 & LG & \\
PCG093956+124037 & 09 39 56.17 & +12 40 37.70 & 4 & 54215$\pm$154 & A & \\
PCG094035+113147 & 09 40 35.69 & +11 31 47.93 & 3 & 22500$\pm$70 & A & \\
PCG094136+121148 & 09 41 36.76 & +12 11 48.84 & 4 & - & C & \\
PCG094321+122625 & 09 43 21.08 & +12 26 25.87 & 4 & 45540$\pm$114 & CP& \\
PCG094756+073010 & 09 47 56.87 & +07 30 10.19 & 4 & 37639$\pm$101 & B & \\
PCG095052+050403 & 09 50 52.18 & +05 04 03.58 & 4 & - & C & \\
PCG095118+122920 & 09 51 18.13 & +12 29 20.04 & 4 & - & C & \\
PCG095507+093520 & 09 55 07.57 & +09 35 20.58 & 4 & 43137$\pm$70 & B &\\
PCG095527+034508 & 09 55 27.22 & +03 45 08.35 & 3 & 27259$\pm$57 & CP& \\
PCG100102-001342 & 10 01 02.72 & -00 13 43.00 & 4 & 27503$\pm$66 & B &\\
PCG100237+063626 & 10 02 37.37 & +06 36 26.32 & 3 & 22750$\pm$60 & A &\\
PCG100355+190454 & 10 03 55.26 & +19 04 54.66 & 4 & 32321$\pm$72 & A & \\
PCG100644+112806 & 10 06 44.41 & +11 28 06.74 & 3 & 46028$\pm$100 & PI& \\
PCG100837+171547 & 10 08 37.84 & +17 15 47.59 & 6 & 37101$\pm$60 & B &
Abell 934\\
PCG101053+034612 & 10 10 53.65 & +03 46 12.90 & 4 & - & C & \\
PCG101113+084127 & 10 11 13.40 & +08 41 27.24 & 3 & 29220$\pm$66 & B &\\
PCG101241-010609 & 10 12 41.24 & -01 06 09.61 & 6 & 29231$\pm$66 & A &\\
PCG101328-005522 & 10 13 28.73 & -00 55 22.01 & 4 & 13263$\pm$62 & B &
Abell 957\\
PCG101345+194541 & 10 13 45.0 & +19 45 41 & 7 & 33465$\pm$89 & B &\\
PCG102512+091835 & 10 25 12.40 & +09 18 35.79 & 3 & 42663$\pm$92 & A &\\
PCG103308+090210 & 10 33 08.01 & +09 02 10.28 & 4 & 67724$\pm$156 & B
& \\
PCG103901+051000 & 10 39 01.83 & +05 10 0.98 & 4 & - & C & \\
PCG103959+274947 & 10 39 59.0 & +27 49 47.0 & 4 & 29876$\pm$74 & A & \\
PCG104215+035811 & 10 42 15.81 & +03 58 11.57 & 4 & - & C & \\
PCG104418+024814 & 10 44 18.96 & +02 48 14.44 & 4 & - & C & \\
PCG104530+202701 & 10 45 30.62 & +20 27 01.84 & 4 & 39051$\pm$99 & B &\\
PCG104538+175827 & 10 45 38.53 & +17 58 27.01 & 4 & - & C & \\
PCG104841+221312 & 10 48 41.98 & +22 13 11.50 & 4 & 13591$\pm$43 & B &
Abell 1100\\
PCG105400+113327 & 10 54 00.74 & +11 33 27.04 & 3 & 45464$\pm$94 & B &\\
PCG110907+022442 & 11 09 07.96 & +02 24 42.01 & 4 & 40315$\pm$92 & A &\\
PCG110941+203320 & 11 09 41.21 & +20 33 20.45 & 3 & 41755$\pm$96 & B &\\
PCG111250+132815 & 11 12 50.41 & +13 28 15.56 & 3 & 50515$\pm$129 & B
& Abell 1201\\
PCG111605+042937 & 11 16 05.89 & +04 29 37.64 & 3 & 33007$\pm$89 & B &\\
PCG111728+074639 & 11 17 28.53 & +07 46 39.72 & 3 & 47379$\pm$129 & B &\\
PCG112051+074439 & 11 20 51.84 & +07 44 39.84 & 4 & - & C & \\
PCG114233+140738 & 11 42 33.12 & +14 07 38.60 & 3 & 37442$\pm$99 & A &\\
PCG114333+215356 & 11 43 33.91 & +21 53 56.72 & 4 & 39661$\pm$100 & A &\\
PCG115606+021907 & 11 56 06.14 & +02 19 07.14 & 4 & - & C &\\
PCG115610+031802 & 11 56 10.09 & +03 18 02.16 & 3 & 21106$\pm$54 & A &\\
PCG120628+081723 & 12 06 28.60 & +08 17 23.42 & 5 & 44385$\pm$92 & B&\\
PCG121157+134421 & 12 11 57.92 & +13 44 21.41 & 4 & - & C &\\
PCG121252+223519 & 12 12 52.51 & +22 35 19.89 & 3 & 25618$\pm$69 & A &\\
PCG121346+072712 & 12 13 46.90 & +07 27 12.85 & 3 & 41091$\pm$80 & B &\\
PCG121359+015956 & 12 13 59.65 & +01 59 56.94 & 4 & - & C & \\
PCG121516+153400 & 12 15 16.01 & +15 34 00.09 & 4 & - & C & \\  
PCG121738+121833 & 12 17 38.05 & +12 18 33.16 & 4 & 280141$\pm$66 & LG &\\
PCG121740+033933 & 12 17 40.61 & +03 39 33.59 & 4 & 23999$\pm$53 & B &\\
PCG122157+080524 & 12 21 57.97 & +08 05 24.93 & 5 & 21541$\pm$54 & B &\\
PCG122222+113923 & 12 22 22.05 & +11 39 23.26 & 4 & 41486$\pm$91 & A &\\
PCG122850-010938 & 12 28 50.96 & -01 09 38.66 & 4 & 34500$\pm$79 & A &\\
PCG122905+083949 & 12 29 05.87 & +08 39 49.43 & 4 & 26699$\pm$70 & B &\\
PCG123437+044539 & 12 34 37.40 & +04 45 40.00 & 4 & - & C & \\
PCG123512+014705 & 12 35 12.23 & +01 47 05.17 & 4 & 24106$\pm$70 & B &
Abell 1564\\
PCG125835+062246 & 12 58 35.16 & +06 22 46.78 & 4 & 24480$\pm$60 & A &\\
PCG130157+191511 & 13 01 57.0 & +19 15 11 & 3 & 23888$\pm$66 & A &\\
PCG130257+053112 & 13 02 57.18 & +05 31 12.47 & 4 & 20813$\pm$18 & CH & \\
PCG130308-022207 & 13 03 08.52 & -02 22 07.97 & 3 & 25342$\pm$54 & B &
Abell 1663\\
PCG130732+074024 & 13 07 32.35 & +07 40 24.60 & 3 & 27865$\pm$63 & A &\\
PCG130926+155358 & 13 09 26.90 & +15 53 58.63 & 4 & 44657$\pm$91 & LG &\\
PCG131132-011944 & 13 11 32.04 & -01 19 44.11 & 4 & - & C & \\
PCG131211+071828 & 13 12 11.56 & +07 18 28.26 & 4 & 27966$\pm$61 & A &\\
PCG131725-014820 & 13 17 25.13 & -01 48 20.59 & 4 & 35261$\pm$87 & A &\\
PCG131730-031041 & 13 17 30.35 & -03 10 41.20 & 4 & - & C & \\
PCG132619+060709 & 13 26 19.11 & +06 07 9.16 & 4 & 25126$\pm$62 & A &\\
PCG132826+012636 & 13 28 26.88 & +01 26 36.74 & 4 & 23437$\pm$53 & B &\\
PCG133042-003302 & 13 30 42.44 & -00 33 02.63 & 3 & 50625$\pm$118 & A &\\
PCG135215+123401 & 13 52 15.45 & +12 33 59.83 & 4 & 43101$\pm$95 & B &
infalling?\\
PCG135456+070521 & 13 54 56.85 & +07 05 21.55 & 4 & 35152$\pm$84 & A &\\
PCG140026+053457 & 14 00 26.50 & +05 34 57.65 & 4 & - & C & \\
PCG140430+102224 & 14 04 30.54 & +10 22 24.99 & 3 & 30494$\pm$63 & A &\\
PCG141129+093748 & 14 11 29.54 & +09 37 48.79 & 4 & 32737$\pm$86 & B &\\
PCG143511+081815 & 14 35 11.69 & +08 18 15.30 & 4 & - & C & \\
PCG143741+185627 & 14 37 41.22 & +18 56 27.06 & 4 & - & C & \\
PCG145239+275905 & 14 52 39.07 & +27 58 50.42 & 3 & 37975$\pm$47 & B &
Abell 1984, infalling group?\\
PCG145853-014235 & 14 58 53.22 & -01 42 35.21 & 4 & - & C & \\
PCG150457+070527 & 15 04 57.59 & +07 05 27.64 & 5 & 27716$\pm$63 & A &\\
PCG150513+134944 & 15 05 13.40 & +13 49 44.72 & 3 & 33271$\pm$82 & A &\\
PCG150708+074838 & 15 07 09.00 & +07 48 38.38 & 4 & - & C & \\
PCG151037+061618 & 15 10 37.26 & +06 16 18.77 & 4 & 51739$\pm$127 & B &\\
PCG151057+031443 & 15 10 57.94 & +03 14 43.33 & 4 & 52283$\pm$110 & LG &\\
PCG151329+025509 & 15 13 29.34 & +02 55 09.26 & 3 & 40327$\pm$98 & B &\\
PCG151340+190714 & 15 13 40.07 & +19 07 14.12 & 4 & - & C & \\
PCG151624+025757 & 15 16 24.76 & +02 57 57.46 & 5 & 33987$\pm$85 & B &\\
PCG151833-013726 & 15 18 33.33 & -01 37 26.58 & 3 & 18790$\pm$123 & A &\\
PCG153046+123131 & 15 30 46.24 & +12 31 31.26 & 4 & 39424$\pm$90 & CP &\\
PCG153147+012457 & 15 31 47.82 & +01 24 57.46 & 4 &  - & C & \\
PCG153234+021221 & 15 32 34.59 & +02 12 21.10 & 4 & 43016$\pm$105 & A &\\
PCG153259+001659 & 15 32 59.95 & +00 16 59.12 & 5 & 24424$\pm$60 & A &\\
PCG154114+034610 & 15 41 14.46 & +03 46 10.42 & 4 &  - & C & \\
PCG154629+005120 & 15 46 29.97 & +00 51 20.23 & 4 &  - & C & \\
PCG154802+030416 & 15 48 02.57 & +03 04 16.54 & 4 & 41682$\pm$64 & A &\\
PCG154930+275637 & 15 49 30.29 & +27 56 41.79 & 4 &  - & C & \\
PCG155024+071836 & 15 50 24.83 & +07 18 36.25 & 4 & 30422$\pm$69 & A &\\
PCG155341+103913 & 15 53 41.51 & +10 39 13.07 & 3 & 56469$\pm$143 & PI &\\
PCG160327+080050 & 16 03 27.25 & +08 00 50.76 & 4 &  - & C & \\
PCG161009+201350 & 16 10 09.27 & +20 13 50.66 & 4 & 45112$\pm$106 & CH &\\
PCG161747+204145 & 16 17 47.86 & +20 41 45.74 & 4 &  - & C & \\
PCG161754+275827 & 16 17 54.0 & +27 58 27 &  4 & 37811$\pm$107 & A &\\
PCG162259+174703 & 16 22 59.25 & +17 47 3.73 & 3 & 34061$\pm$91 & A &\\
PCG170458+281834 & 17 04 57 &  +28 18 3  & 4 &  - & C & \\
PCG220719-020723 & 22 07 19.72 & -02 07 23.59 & 4 &  - & C & \\
PCG220748-004159 & 22 07 48.64 & -00 41 59.17 & 4 & 32800$\pm$83 & A &\\
PCG220929+012412 & 22 09 29.11 & +01 24 12.64 & 3 & 25545$\pm$59 & A &\\
PCG221414+002203 & 22 14 14.50 & +00 22 3.22 & 5 & 38111$\pm$86 & A &\\
PCG221442+012823 & 22 14 42.20 & +01 28 23.52 & 4 & 54847$\pm$127 & CH &\\
PCG221755-013227 & 22 17 55.91 & -01 32 27.89 & 4 & 30039$\pm$63 & A &\\
PCG222037+053440 & 22 20 37.15 & +05 34 40.58 & 4 &  - & C & \\
PCG222111-010504 & 22 21 11.76 & -01 05 04.88 & 4 &  - & C & \\
PCG222121+002743 & 22 21 21.97 & +00 27 43.24 & 4 &  - & C & \\
PCG222450+071501 & 22 24 50.90 & +07 15 01.91 & 4 &  - & C & \\
PCG222633+051207 & 22 26 33.57 & +05 12 07.02 & 4 & 30634$\pm$71 & CH &\\
PCG222900+003815 & 22 29 00.57 & +00 38 15.36 &  4 &  - & C & \\
PCG223216+034245 & 22 32 16.30 & +03 42 45.61 & 5 & 17602$\pm$45 & A &\\
PCG223922-005611 & 22 39 22.24 & -00 56 11.26 & 3 & 40177$\pm$101 & A &\\
PCG224346+120335 & 22 43 46.88 & +12 03 35.50 & 4 & 24064$\pm$64 & A &\\
PCG224713+020337 & 22 47 13.50 & +02 03 37.84 & 4 &  - & C & \\
PCG224931-011755 & 22 49 31.55 & -01 17 55.07 & 4 &  - & C & \\
PCG225556+094906 & 22 55 56.54 & +09 49 06.31 & 4 &  - & C & \\
PCG225807+011101 & 22 58 07.05 & +01 11 01.21 & 4 & 30813$\pm$79 & LG &\\
PCG231910-022709 & 23 19 10.28 & -02 27 09.90 & 6 & 32709$\pm$75 & B &
Abell 2571\\
PCG233446+003743 & 23 34 46.19 & +00 37 43.46 & 4 &  - & C & \\
PCG234100+000450 & 23 41 00.07 & +00 04 50.30 & 5 & 55903$\pm$134 & B
& Abell 2644\\
PCG235439+032308 & 23 54 39.98 & +03 23 08.63 & 5 & 26701$\pm$65 & A &\\
\hline\hline
\end{longtable}

\setcounter{table}{1}
\begin{longtable}{c c c c c c c c}
\caption{\label{group_prop} Group dynamical
  properties. $H_\mathrm{o}t_\mathrm{c}$ is expressed in logarithmic
  units, while all the other quantities are indicated in their natural
  units.}\\
\hline\hline
Group name   &   R        & Scale                   & $\sigma_{r}$ & $H_\mathrm{o}t_\mathrm{c}$ & M                    & L                          & M/L   \\ 
                      & arcmin & (kpc/$\arcmin$) & (\kms)          & &$M_\mathrm{\sun}$    & $L_\mathrm{\sun}$ \\ \hline

PCG001108+054449 & 0.2220 & 160.22 &  89.01 &  -1.489	& 9.2$\times10^{11}$ &	1.4 $\times 10^{11}$	  & 7 \\
PCG011206+042617 & 0.5100 & 126.77 & 238.75 &  -1.746 	& 1.1 $\times10^{13}$ &	7.9 $\times 10^{10}$	& 139\\
PCG015254-001033 & 0.4820 &  96.24 & 207.52 &  -1.874 	& 6.5 $\times 10^{12}$&	7.4 $\times 10^{10}$	 & 88\\
PCG025234+111647 & 0.3430 & 133.88 & 250.66 &  -1.951 	& 9.4 $\times 10^{12}$&	8.8 $\times 10^{10}$	& 107\\
PCG025903+100636 & 0.6670 & 135.18 & 180.76 &  -1.461 	& 8.8 $\times10^{12}$ &	1.1 $\times 10^{11}$	& 79\\
PCG030301+052405 & 0.3246 & 117.35 & 189.62 &  -1.854 	& 4.1 $\times10^{12}$ & 8.9 $\times 10^{10}$	 & 46\\
PCG030352+084700 & 0.5450 & 105.46 & 288.53 &  -1.927 	& 1.6 $\times10^{13}$ & 1.5 $\times 10^{11}$	& 103\\
PCG031139+072404 & 0.3000 & 161.01 & 213.35 &  -1.834 	& 7.2 $\times 10^{12}$	& 2.3 $\times 10^{11}$	 & 32\\
PCG091524+213038 & 0.3240 & 148.82 & 430.21 &  -2.200 	& 3.2 $\times 10^{13}$	& 1.1 $\times 10^{11}$	& 290\\
PCG093220+171954 & 0.5318 & 155.30 & 1150.41 &  -2.362 	& 3.3 $\times 10^{14}$	& 1.0 $\times 10^{11}$	& 3267\\
PCG093226+094339 & 0.4400 &  91.30 & 489.72 &  -2.328 	& 3.1 $\times 10^{13}$	& 5.5 $\times 10^{10}$	& 572\\
PCG093310+092639 & 1.8378 & 154.16 & 103.17 &  -0.869 	& 9.0 $\times 10^{12}$	& 1.3 $\times 10^{11}$	& 71\\
PCG093956+124037 & 0.4910 & 191.07 & 198.93 &  -1.384 	& 1.2 $\times 10^{13}$	& 1.9 $\times 10^{11}$	& 64\\
PCG094035+113147 & 0.6430 &  90.21 & 313.56 &  -1.939 	& 1.7 $\times 10^{13}$	& 3.9 $\times 10^{10}$	& 432\\
PCG094321+122625 & 0.2820 & 165.70 & 487.91 &  -2.252 	& 3.6 $\times 10^{13}$	& 1.6 $\times 10^{11}$	& 228\\
PCG094756+073010 & 0.2290 & 142.45 & 609.66 &  -2.510 	& 4.0 $\times 10^{13}$	& 1.4 $\times 10^{11}$	& 292\\
PCG095507+093520 & 0.2900 & 159.89 & 601.13 &  -2.352 	& 5.5 $\times 10^{13}$	& 2.4 $\times 10^{11}$	& 230\\
PCG095527+034508 & 0.4504 & 106.30 & 1049.63 &  -2.559 	& 1.6 $\times 10^{14}$	& 7.6 $\times 10^{10}$	& 2063\\
PCG100102-001342 & 0.3830 & 108.18 &  89.39 &  -1.376 	& 1.1 $\times 10^{12}$	& 1.1 $\times 10^{11}$	&  10\\
PCG100237+063626 & 0.4430 &  90.20 & 133.45 &  -1.635 	& 2.1 $\times 10^{12}$	& 4.0 $\times 10^{10}$	 & 54\\
PCG100355+190454 & 0.4250 & 123.54 & 141.45 &  -1.601 	& 3.4 $\times 10^{12}$	& 1.0 $\times 10^{11}$	& 34\\
PCG100644+112806 & 0.6062 & 167.15 & 1236.92 &  -2.304 	& 4.6 $\times 10^{14}$	& 1.3 $\times 10^{11}$	& 3638\\
PCG100837+171547 & 0.5140 & 139.34 & 523.38 &  -2.125 	& 7.5 $\times 10^{13}$	& 1.5 $\times 10^{11}$	& 484\\
PCG101113+084127 & 0.7930 & 114.15 &  95.67 &  -1.070 	& 2.5 $\times 10^{12}$	& 1.4 $\times 10^{11}$	& 17\\
PCG101241-010609 & 0.5010 & 114.15 & 160.62 &  -1.676 	& 5.6 $\times 10^{12}$	& 1.2 $\times 10^{11}$	& 47\\
PCG101328-005522 & 0.3390 &  55.84 & 518.93 &  -2.678 	& 1.7 $\times 10^{13}$	& 1.7 $\times 10^{10}$	& 991\\
PCG101345+194541 & 0.5100 & 127.45 & 262.65 &  -1.859 	& 1.8 $\times 10^{13}$	& 1.6 $\times 10^{11}$	& 113\\
PCG102512+091835 & 0.3890 & 156.88 &  59.15 &  -1.294 	& 6.4$\times 10^{11}$	& 9.9 $\times 10^{10}$	  & 6\\
PCG103308+090210 & 0.2118 & 229.50 & 343.36 &  -2.048 	& 1.9 $\times 10^{13}$	& 1.7 $\times 10^{11}$	& 113\\
PCG103959+274947 & 0.2540 & 116.44 & 311.81 &  -2.253 	& 9.4 $\times 10^{12}$	& 6.3 $\times 10^{10}$	& 148\\
PCG104530+202701 & 0.2700 & 147.00 & 470.50 &  -2.309 	& 2.9 $\times 10^{13}$	& 1.7 $\times 10^{11}$	& 170\\
PCG104841+221312 & 0.5650 &  55.88 & 506.40 &  -2.448 	& 2.6 $\times 10^{13}$	& 3.5 $\times 10^{10}$	& 750\\
PCG105400+113327 & 0.3820 & 167.02 & 263.88 &  -1.793 	& 1.3 $\times 10^{13}$	& 2.2 $\times 10^{11}$	& 61\\
PCG110907+022442 & 0.5000 & 149.58 & 263.86 &  -1.756 	& 1.7 $\times 10^{13}$	& 8.7 $\times 10^{10}$	& 194\\
PCG110941+203320 & 0.4010 & 155.58 & 239.65 &  -1.758 	& 1.1 $\times 10^{13}$	& 7.9 $\times 10^{10}$	& 135\\
PCG111250+132815 & 0.3200 & 182.91 & 291.08 &  -1.866 	& 1.5 $\times 10^{13}$	& 1.1 $\times 10^{11}$	& 132\\
PCG111605+042937 & 0.3200 & 125.85 & 137.65 &  -1.622 	& 2.3 $\times 10^{12}$	& 7.1 $\times 10^{10}$	& 32\\
PCG111728+074639 & 0.4100 & 172.84 & 310.51 &  -1.808 	& 2.0 $\times 10^{13}$	& 1.4 $\times 10^{11}$	& 149\\
PCG114233+140738 & 0.2150 & 140.41 & 361.96 &  -2.279 	& 1.2 $\times 10^{13}$	& 9.4 $\times 10^{10}$	& 126\\
PCG114333+215356 & 0.2750 & 147.49 &  93.90 &  -1.672 	& 1.2 $\times 10^{12}$	& 1.7 $\times 10^{11}$	&  7\\
PCG115610+031802 & 0.3536 &  85.13 & 297.91 &  -2.204 	& 8.0 $\times 10^{12}$	& 2.1 $\times 10^{10}$	& 372\\
PCG120628+081723 & 0.4000 & 163.71 & 571.78 &  -2.194 	& 7.6 $\times 10^{13}$	& 1.4 $\times 10^{11}$	& 523\\
PCG121252+223519 & 0.2860 & 101.49 & 143.95 &  -1.682 	& 1.8 $\times 10^{12}$	& 9.6 $\times 10^{10}$	& 19\\
PCG121346+072712 & 0.7160 & 152.04 & 264.03 &  -1.579 	& 2.3 $\times 10^{13}$	& 1.4 $\times 10^{11}$	& 162\\
PCG121738+121833 & 0.3560 & 108.90 & 438.88 &  -2.293 	& 2.4 $\times 10^{13}$	& 7.3 $\times 10^{10}$	& 336\\
PCG121740+033933 & 0.4662 &  95.67 & 961.25 &  -2.578 	& 1.3 $\times 10^{14}$	& 5.4 $\times 10^{10}$	& 2476\\
PCG122157+080524 & 0.5920 &  86.68 & 130.88 &  -1.620 	& 3.1 $\times 10^{12}$	& 7.4 $\times 10^{10}$	& 42\\
PCG122222+113923 & 0.3740 & 153.25 & 117.31 &  -1.361 	& 2.6 $\times 10^{12}$	& 1.1 $\times 10^{11}$	& 24\\
PCG122850-010938 & 0.3450 & 130.82 & 194.04 &  -1.832 	& 5.5 $\times 10^{12}$	& 9.2 $\times 10^{10}$	& 60\\
PCG122905+083949 & 0.3360 & 104.30 & 499.88 &  -2.394 	& 2.9 $\times 10^{13}$	& 5.4 $\times 10^{10}$	& 534\\
PCG123512+014705 & 0.5540 &  96.10 & 446.58 &  -2.162 	& 3.5 $\times 10^{13}$	& 5.4 $\times 10^{10}$	& 641\\
PCG125835+062246 & 0.4350 &  96.45 & 194.89 &  -1.888 	& 5.2 $\times 10^{12}$	& 8.1 $\times 10^{10}$	& 65\\
PCG130157+191511 & 0.3350 &  94.30 &  77.50 &  -1.375 		& 5.7 $\times 10^{11}$	& 6.9 $\times 10^{10}$	  & 8\\
PCG130257+053112 & 0.5400 &  84.02 &  79.13 &  -1.451 		& 9.3 $\times 10^{11}$	& 9.2 $\times 10^{10}$	 & 10\\
PCG130308-022207 & 0.5140 &  99.55 & 526.07 &  -2.229 	& 4.2 $\times 10^{13}$	& 5.0 $\times 10^{10}$	& 849\\
PCG130732+074024 & 0.6150 & 109.45 & 212.54 &  -1.693 	& 9.1 $\times 10^{12}$	& 6.4 $\times 10^{10}$	& 141\\
PCG130926+155358 & 0.6820 & 164.63 & 341.39 &  -1.713 	& 4.3 $\times 10^{13}$	& 1.5 $\times 10^{11}$	& 285\\
PCG131211+071828 & 0.3350 & 108.69 & 247.55 &  -2.062 	& 7.3 $\times 10^{12}$	& 9.6 $\times 10^{10}$	& 76\\
PCG131725-014820 & 0.2750 & 133.28 & 318.96 &  -2.167 	& 1.2 $\times 10^{13}$	& 1.3 $\times 10^{11}$	& 97\\
PCG132619+060709 & 0.3860 &  98.79 & 111.10 &  -1.629 	& 1.5 $\times 10^{12}$	& 5.1 $\times 10^{10}$	& 30\\
PCG132826+012636 & 0.5570 &  93.60 & 102.77 &  -1.460 	& 1.8 $\times 10^{12}$	& 8.3 $\times 10^{10}$	 & 21\\
PCG133042-003302 & 0.3380 & 182.56 & 214.74 &  -1.671 	& 8.5 $\times 10^{12}$	& 1.2 $\times 10^{11}$	& 69\\
PCG135215+123401 & 0.1930 & 159.80 & 278.10 &  -2.172 	& 7.8 $\times 10^{12}$	& 1.2 $\times 10^{11}$	& 65\\
PCG135456+070521 & 0.3520 & 134.29 & 298.32 &  -2.025 	& 1.4 $\times 10^{13}$	& 7.5 $\times 10^{10}$	& 184\\
PCG140430+102224 & 0.5122 & 117.35 &  73.57 &  -1.395 	& 9.7 $\times 10^{11}$	& 5.3 $\times 10^{10}$	 & 18\\
PCG141129+093748 & 0.2770 & 124.95 & 371.19 &  -2.261 	& 1.6 $\times 10^{13}$	& 5.9 $\times 10^{10}$	& 262\\
PCG145239+275905 & 0.4223 & 143.52 & 223.51 &  -1.766 	& 9.1 $\times 10^{12}$	& 4.2 $\times 10^{10}$	& 216\\
PCG150457+070527 & 0.5240 & 108.91 & 211.25 &  -1.805 	& 9.0 $\times 10^{12}$	& 1.2 $\times 10^{11}$	& 75\\
PCG150513+134944 & 0.4350 & 127.98 & 166.45 &  -1.626 	& 4.6 $\times 10^{12}$	& 1.0 $\times 10^{11}$	& 44\\
PCG151037+061618 & 0.5220 & 183.80 & 384.28 &  -1.797 	& 4.6 $\times 10^{13}$	& 1.1 $\times 10^{11}$	& 415\\
PCG151057+031443 & 0.2660 & 185.59 & 514.35 &  -2.253 	& 4.3 $\times 10^{13}$	& 1.7 $\times 10^{11}$	& 250\\
PCG151329+025509 & 0.8560 & 149.58 & 153.25 &  -1.149 	& 9.0 $\times 10^{12}$	& 2.0 $\times 10^{11}$	& 45\\
PCG151624+025757 & 0.2820 & 130.39 & 393.18 &  -2.277 	& 2.0 $\times 10^{13}$	& 1.2 $\times 10^{11}$	& 175\\
PCG151833-013726 & 0.5230 &  76.65 &  14.93 &  -0.879 		& 2.7 $\times 10^{10}$	& 7.0 $\times 10^{10}$	  & 0.382\\
PCG153046+123131 & 0.4574 & 148.16 & 816.11 &  -2.323 	& 1.5 $\times 10^{14}$	& 1.7 $\times 10^{11}$	& 879\\
PCG153234+021221 & 0.3350 & 158.00 & 213.16 &  -1.774 	& 7.9 $\times 10^{12}$	& 1.5 $\times 10^{11}$	& 51\\
PCG153259+001659 & 0.6440 &  96.24 & 134.69 &  -1.524 	& 4.0 $\times 10^{12}$	& 8.5 $\times 10^{10}$	& 47\\
PCG154802+030416 & 0.2789 & 155.30 & 132.87 &  -1.665 	& 2.5 $\times 10^{12}$	& 1.0 $\times 10^{11}$	& 25\\
PCG155024+071836 & 0.4560 & 117.16 & 262.62 &  -1.922 	& 1.2 $\times 10^{13}$	& 8.4 $\times 10^{10}$	& 143\\
PCG155341+103913 & 0.5008 & 197.38 & 708.57 &  -2.066 	& 1.5 $\times 10^{14}$	& 1.5 $\times 10^{11}$	& 993\\
PCG161009+201350 & 0.2190 & 166.01 & 473.26 &  -2.348 	& 2.7 $\times 10^{13}$	& 1.3 $\times 10^{11}$	& 201\\
PCG161754+275827 & 0.2860 & 141.57 & 169.78 &  -1.685 	& 3.8 $\times 10^{12}$	& 8.1 $\times 10^{10}$	& 47\\
PCG162259+174703 & 0.3940 & 129.34 & 224.11 &  -1.832 	& 7.7 $\times 10^{12}$	& 1.2 $\times 10^{11}$	& 65\\
PCG220748-004159 & 0.4640 & 126.37 & 103.90 &  -1.251 	& 2.1 $\times 10^{12}$	& 1.9 $\times 10^{11}$	& 11\\
PCG220929+012412 & 0.2710 & 100.29 &  75.55 &  -1.310 	& 4.6 $\times 10^{11}$	& 2.8 $\times 10^{10}$	& 17\\
PCG221414+002203 & 0.3620 & 142.53 & 172.98 &  -1.719 	& 5.5 $\times 10^{12}$	& 8.9 $\times 10^{10}$	& 61\\
PCG221442+012823 & 0.3587 & 194.68 & 363.99 &  -1.937 	& 3.0 $\times 10^{13}$	& 1.4 $\times 10^{11}$	& 219\\
PCG221755-013227 & 0.3390 & 115.83 & 189.22 &  -1.901 	& 4.6 $\times 10^{12}$	& 9.1 $\times 10^{10}$	& 50\\
PCG222633+051207 & 0.4630 & 119.02 & 715.00 &  -2.356 	& 9.2 $\times 10^{13}$	& 1.3 $\times 10^{11}$	& 714\\
PCG223216+034245 & 0.6320 &  71.96 & 341.04 &  -2.130 	& 1.9 $\times 10^{13}$	& 1.9 $\times 10^{11}$	& 99\\
PCG223922-005611 & 0.5110 & 150.55 &  96.63 &  -1.406 	& 2.1 $\times 10^{12}$	& 6.0 $\times 10^{10}$	& 35\\
PCG224346+120335 & 0.4710 &  94.95 & 172.56 &  -1.795 	& 4.3 $\times 10^{12}$	& 5.6 $\times 10^{10}$	& 77\\
PCG225807+011101 & 0.3260 & 118.47 & 542.84 &  -2.387 	& 3.7 $\times 10^{13}$	& 1.1 $\times 10^{11}$	& 332\\
PCG231910-022709 & 0.5930 & 124.28 & 194.59 &  -1.645 	& 1.1 $\times 10^{13}$	& 1.0 $\times 10^{11}$	& 103\\
PCG234100+000450 & 0.2580 & 197.66 & 279.88 &  -1.946 	& 1.4 $\times 10^{13}$	& 3.2 $\times 10^{11}$	& 44\\
PCG235439+032308 & 0.5140 & 104.30 & 178.23 &  -1.746 	& 6.0 $\times10^{12}$	& 9.4 $\times 10^{10}$	& 64\\
\hline\hline
\end{longtable} 

\Online
 We present here a set of figures representative of {\it class A},
 {\it class B} groups.

\setcounter{figure}{7}
\begin{figure*}
\includegraphics{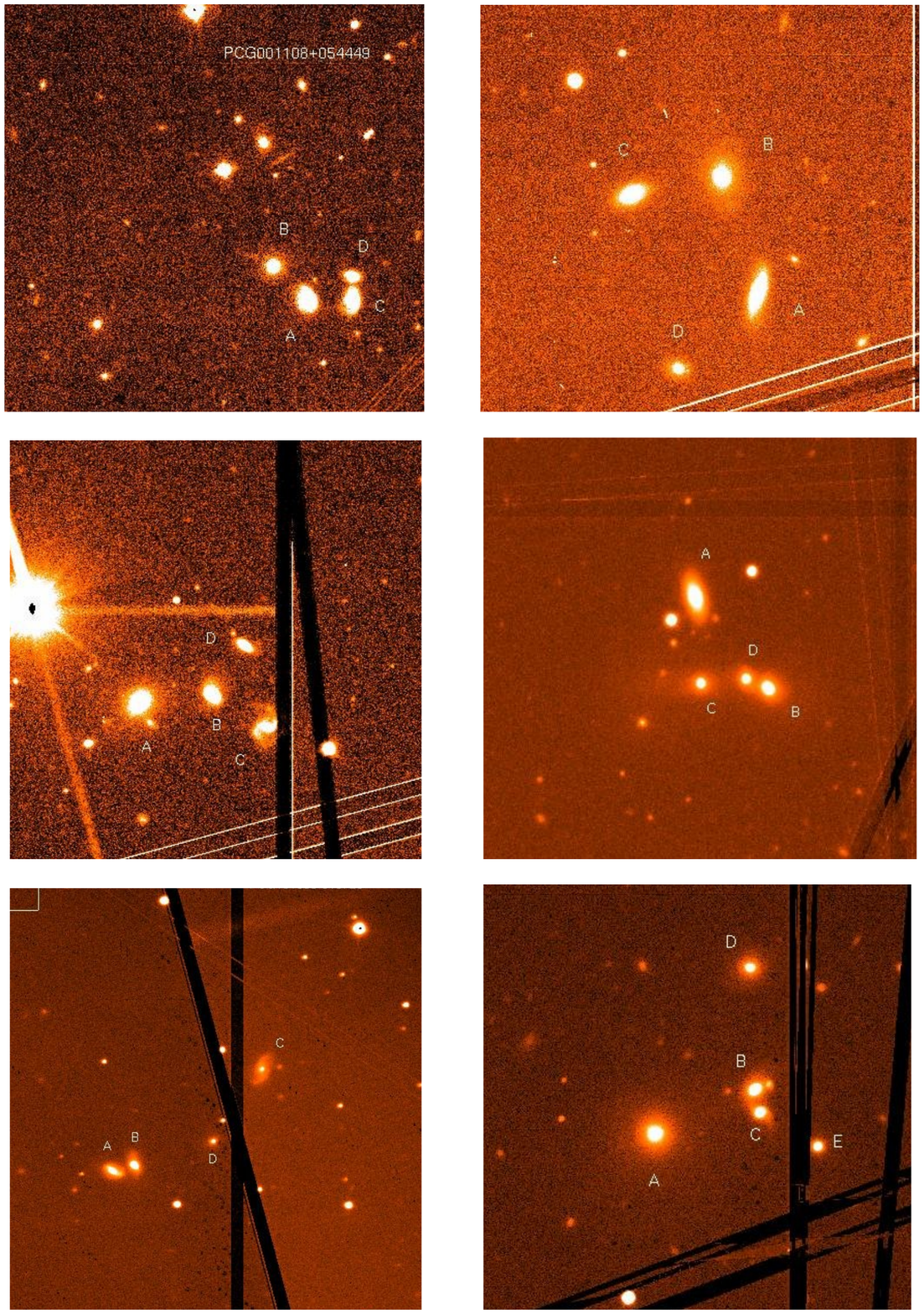}
\caption{Stacked R band acquisition images for a sample of {\it class
    A} confirmed compact groups. From top left to bottom right, in
  increasing right ascension order: PCG001108+054449,
  PCG015254-001033, PCG102512+091835, PCG114233+140738,
  PCG151833-013726, and PCG235439+032308. The black lines visible in
  some
images are the gap between the two CCDs on the EMMI red mosaic. In all
cases
the whole group was in one chip, but imaged at different rotation
angles, to position accurately the slit. Exposure times vary from 180s
to 450s. In all images N is
up and E is left.}
\end{figure*}

\afterpage{\clearpage}

\setcounter{figure}{8}
\begin{figure*}
\includegraphics{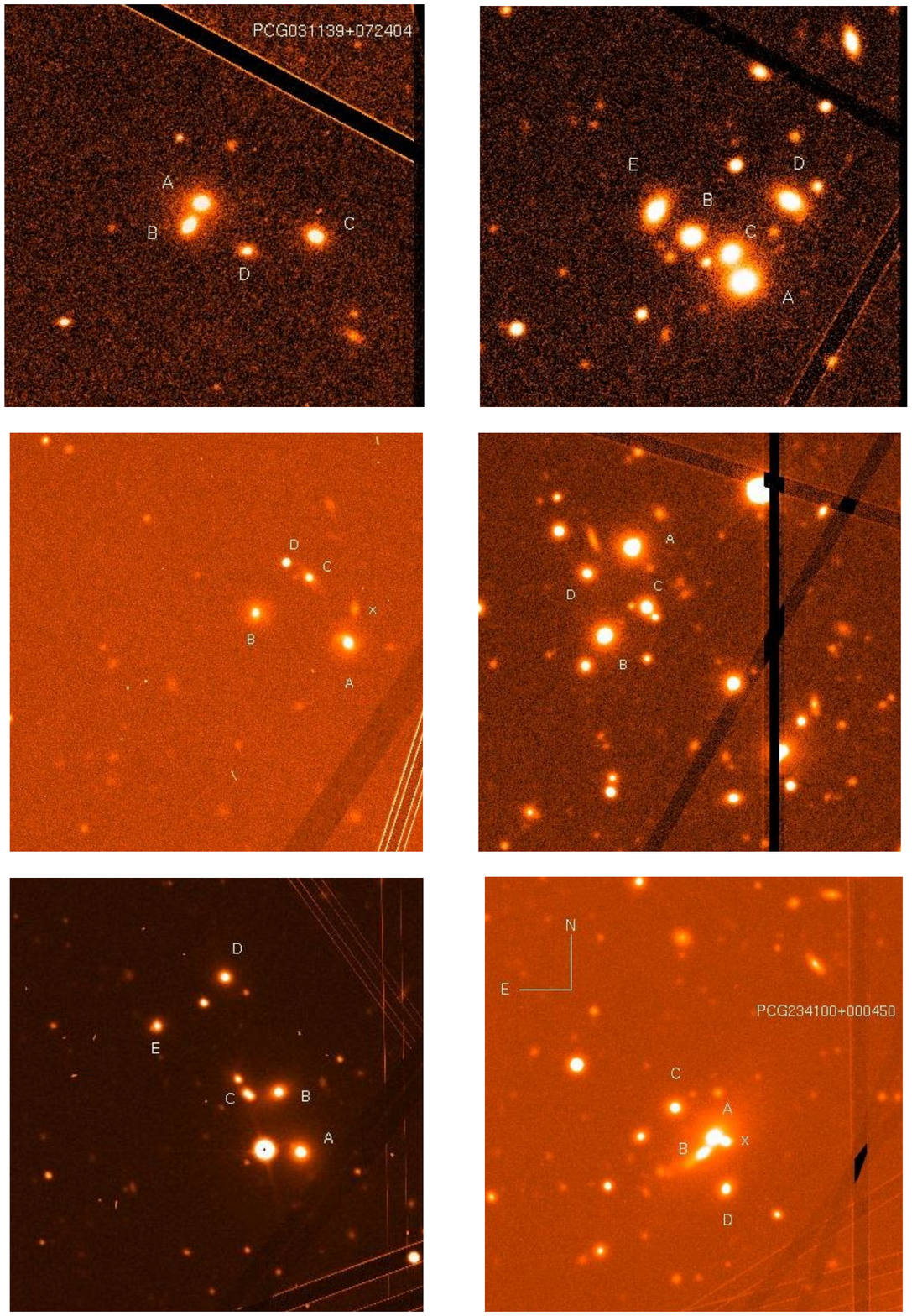}
\caption{Same for Fig.8, but for class B groups. From top left to bottom right, in
  increasing right ascension order: PCG031139+072404,
  PCG091524+213038, PCG105400+113327, PCG111250+132815,
  PCG151037+061618, and PCG234100+000450.}
\end{figure*}

\afterpage{\clearpage}


\begin{thebibliography}{}

\bibitem[2009]{aba09} Abazajian, N.K, and the SDSS team, 2009, ApJS, 182, 543

\bibitem[2005]{alla05} Allam, S.S.; Tucker, D.L.; Allyn Smyth, J.,
  2005, AJ, 129, 5

\bibitem[2005]{ande05} Andernach, H.; Coziol, R., 2005, ASP, 329, 67

\bibitem[1985]{andersen85} Andersen, J., et al. 1985, A\&AS, 59, 15

\bibitem[1989]{barnes89} Barnes, J.E., Nature, 338, 123

\bibitem[1990]{beers90} Beers, T., Flynn, K., Gebhardt,K., 1990
AJ, 100, 32

\bibitem[2000]{bra00} Bramel, D. A.; Nichol, R. C.; Pope, A. C., 2000,
  ApJ 533, 601

\bibitem[2001]{colb01} Colber, J.W.; Mulchaey, J.S.; Zabludoff, A.,
  2001, AJ, 121, 808

\bibitem[2006]{collo06} Collobert, M.; Sarzi, M.; Davies, R.L.;
  Kuntschner, H.; Colless, M., 2006, MNRAS, 370, 1213

\bibitem[1994]{reina94} de Carvalho, R. R.; Ribeiro, A.; Zepf, S., 1994, ApJS, 93,
  47

\bibitem[2005]{reina05} de Carvalho, R. R.; Goncalves, T. S.; Iovino A.; 
Kohl-Moreira, J. L;, Gal, R. R.; Djorgovski, S.G., 2005, AJ, 130, 425

\bibitem[2003]{} Einasto,M;  Einasto, J.;  Mueller, V., Heinamaki, P.;
  Tucker, D.L., 2003, A\&A 401, 851

\bibitem[2004]{gal04} Gal, R. R.; de Carvalho, R. R.; Odewahn, S. C,
  Djorgovski et al.; 2004, AJ, 128, 3082

\bibitem[1980]{gisl80} Gisler, G.R., 1980, AJ, 85, 623

\bibitem[2000]{Giuri00} Giuricin, G.; Marinoni, C.; Ceriani, L.;
  Pisani, A., 2000, ApJ, 543, 178

\bibitem[2011]{guti11} Gutierrez, C.M., 2011, arXiv:1107.2815v1,
  accepted on ApJL

\bibitem[1985]{heisler85} Heisler, J., Tremaine, S., Bahcall, J., 1985,
ApJ, 298, 8

\bibitem[]{}
Hickson, P. 1982, ApJ, 255, 382

\bibitem[1992]{hickson92} Hickson, P., Mendes de Oliveira, C., Huchra, J. P., 
Palumbo, G. G. C., 1992, ApJ, 399, 353

\bibitem[2003]{iovino03}  Iovino, A.; de Carvalho R.R.; Gal, R. R.; Odewahn, S. C.;
Lopes, P. A. A.; Mahabal A.; Djorgovski, S. G., 2003, AJ, 125, 1660

\bibitem[2003]{jo03} Jones, L. R.; Ponman, T. J.; Horton, A., et
  al. 2003, MNRAS, 343, 627

\bibitem[1985]{} Kent, S.M., 1985, PASP, 97, 165

\bibitem[2007]{koe07} Koester, B.P. et al., 2007, ApJ, 660, 239

\bibitem[1998]{kurtz98} Kurtz, M. J.; Mink, D. J., 1998
PASP, 110, 934

\bibitem[2004]{} Lee, B.C.; Allam, S. S.; Tucker, D.L.;  Annis, J. et
  al., 2004, AJ, 127, 1811

\bibitem[2008]{long08} Longair, M.S.; 2008, Galaxy formation, A\&A library,
  Springer Verlag

\bibitem[20044]{lop04} Lopes, P.A.A.; de Carvalho, R.R.; Gal, R.R.;
  Djorgovski, S.G.;  Odewahn, S.C.; Mahabal, A.A. and R. J. Brunner,
  2004, AJ, 128, 1017

\bibitem[2008]{ma08} Martinez, M.A.; Del Olmo, A; Coziol, R.; Focardi,
  P.; 2008, ApJ, 678, L9

\bibitem[1994]{mdo94} Mendes de Oliveira, C.; Hickson, P., 1994, ApJ,
  427, 684

\bibitem[2005]{mdo05} Mendes de Oliveira, C.;  Coelho, P.; Gonzalez,
  J.J. and Barbuy, B; 2005, AJ, 130, 55

\bibitem[2009]{mc09} McConnachie, A.W.; Patton, D.R.; Ellison, S.L.;
  Simard, L., 2009, MNRAS, 395, 255

\bibitem[1973]{nil73} Nilson, P., 1973, Uppsala general catalogue of
  galaxies

\bibitem[1990]{perea90} Perea, J., del Olmo, A., Moles, M., 1990,
A\&A, 237, 319

\bibitem[2003]{ela03} Pompei E.; Iovino, A., 2003, Ap\&SS, 285, 133
 
\bibitem[2006]{ela} Pompei, E.; de Carvalho, R.R.; Iovino, A., 2006, A\&A, 445, 857;
Paper I

\bibitem[2007]{ela07} Pompei, E.; Dahlem, M.; A. Iovino; 2007, A\&A, 473, 399, 2007

\bibitem[]{}
Ponman, T.J., Bourner, P.D.J., Ebeling, H., Bohringer, H., 1996
MNRAS, 283, 690

\bibitem[1998]{ribe98} Ribeiro, A., de Carvalho, R. R., Capelato, H., Zepf, S. E., 1998
ApJ, 497, 72

\bibitem[2007]{san07} Santos, W.A.; Mendes de Oliveira, C; Laerte
  Sodre, Jr., 2007, AJ, 134, 1551

\bibitem[2010]{tago10} Tago, E.; Saar, E.; Tempel, E.; Einasto, J.;
  Nurmi, P.; Heinamaki, P., 2010, A\&A, 514, 102

\bibitem[1979]{td} Tonry, J.; Davis, M.; 1979, AJ, 84, 1511

\bibitem[2005]{vdk05} van Dokkum, P.,  2005, AJ, 130, 2647

\bibitem[2001]{vm01}Verdes-Mentenegro, L., Yun, M. S., Williams, B. A., et al. 2001,
   A\&A, 377, 812

\bibitem[2007]{ver07} Verley, S.; Odewahn, S.C.; Verdes-Montenegro,
  L.; Leon, S.; Combes, F. et al., 2007, A\&A, 470, 505

\bibitem[1989]{} West, M., 1989, ApJ, 344, 535
 
\bibitem[1991]{wind91}  Windhorst, R. A. et al., 1991
ApJ, 380, 362
 
\end{thebibliography}
\end{document}